\documentclass[12pt,nofootinbib,superscriptaddress]{revtex4}

\usepackage{amssymb,color,paralist,amsmath,epsfig}
\usepackage{graphicx}
\usepackage{comment}
\usepackage{amsfonts}
\usepackage{array}
\usepackage{fancypar}
\usepackage{xcolor}
\usepackage{relsize}

\newcommand{\cG}{{\cal G}}
\newcommand{\cF}{{\cal F}}
\newcommand{\cma}{{\theta_{\mathrm{cm}}}}

\newcommand{\hp}{{\frac{1}{2}}}
\newcommand{\hm}{{-\frac{1}{2}}}

\newcommand{\beq}{\begin{equation}}
\newcommand{\eeq}{\end{equation}}

\newcommand{\ber}{\begin{eqnarray}} 
\newcommand{\eer}{\end{eqnarray}}

\begin{document}

\title{ Subtracted dispersion relation formalism for the two-photon exchange correction to elastic electron-proton scattering: comparison with data }

\author{O. Tomalak}
\affiliation{Institut f\"ur Kernphysik, Johannes Gutenberg Universit\"at, Mainz, Germany}
\affiliation{PRISMA Cluster of Excellence, Johannes Gutenberg-Universit\"at,  Mainz, Germany}
\affiliation{Department of Physics, Taras Shevchenko National University of Kyiv, Kyiv, Ukraine}
\author{M. Vanderhaeghen}
\affiliation{Institut f\"ur Kernphysik, Johannes Gutenberg Universit\"at, Mainz, Germany}
\affiliation{PRISMA Cluster of Excellence, Johannes Gutenberg-Universit\"at,  Mainz, Germany}
\date{\today}

\begin{abstract}
We apply a subtracted dispersion relation formalism with the aim to 
improve predictions for the two-photon exchange corrections to elastic electron-proton scattering observables at finite momentum transfers. 
We study the formalism on the elastic contribution, and make a detailed comparison with existing data for unpolarized cross sections 
as well as polarization transfer observables.
\end{abstract}

\maketitle

\section{Introduction}
\label{sec1}

Lepton scattering within the one-photon exchange approximation is a time honored
tool to access information on the internal structure of hadrons, in particular the distribution of
charge and magnetization within a nucleon. The traditional way to access nucleon form factors (FFs) - the Rosenbluth separation technique, measures the angular dependence of the unpolarized differential cross section for elastic electron-nucleon scattering. Electric and magnetic FFs have 
been measured with this technique, see Refs.~\cite{Bernauer:2010wm, Bernauer:2013tpr} for such recent state-of-the-art measurements, and e.g. Ref.~\cite{Perdrisat:2006hj} for a review of older data. The development of the recoil polarization technique as well as the availability of polarized targets at electron scattering facilities led to the possibility of 
a second method of FF measurements. Such experiments access the ratio $ G_E/G_M $ of electric ($G_E$) to magnetic ($G_M$) FFs directly from the ratio of the transverse to longitudinal nucleon polarizations in elastic electron-nucleon scattering. For squared momentum transfers $ Q^2 $ up to 8.5 $ \mathrm{GeV}^2 $, this ratio has been measured at Jefferson Laboratory (JLab) in a series of experiments~\cite{Jones:1999rz, Gayou:2001qd, Punjabi:2005wq, Puckett:2010ac}, with projects to extend these measurements in the near future at the JLab 12 GeV facility to even larger $Q^2$ values~\cite{Punjabi:2014nea}. It came as a surprise that the two experimental approaches to access nucleon FFs, assuming the single-photon exchange approximation, gave strikingly different results for the FF ratio, for $ Q^2 $ value above 1.0 $ \mathrm{GeV}^2$.  
Two-photon exchange (TPE) processes have been proposed as a plausible solution to resolve this 
puzzle~\cite{Blunden:2003sp, Guichon:2003qm}, see Ref.~\cite{Carlson:2007sp} for a review.
Estimates for TPE processes were studied in a variety of different model calculations, see e.g. 
Refs.~\cite{Blunden:2003sp, Chen:2004tw, Afanasev:2005mp, Blunden:2005ew,  Gorchtein:2006mq, Borisyuk:2008es, Borisyuk:2008db, Kivel:2009eg, Kivel:2012vs, Graczyk:2013pca, Borisyuk:2013hja}, and first phenomenological extractions of TPE observables based on available data were given, see e.g. Refs.~\cite{Arrington:2007ux, Chen:2007ac, Borisyuk:2007re, Belushkin:2007zv, Qattan:2011ke, Guttmann:2010au}. Furthermore,  dedicated experiments to directly measure the TPE observables have been performed in recent years~\cite{Meziane:2010xc, Moteabbed:2013isu}, or are underway~\cite{Gramolin:2011tr, Milner:2013daa}. 

Besides electron scattering experiments, information on the proton size can also be obtained from 
atomic spectroscopy. Theoretical predictions for the hydrogen spectrum within QED are performed to such accuracy that they can be used as a precision tool to extract the proton radius, 
see e.g. Ref.~\cite{Antognini:2013rsa} for a recent work and references therein. It came as a surprise that the recent extractions of the proton charge radius from muonic hydrogen Lamb shift measurements~\cite{Pohl:2010zza, Antognini:1900ns}  are in strong contradiction, by around 7 standard deviations, with the values obtained from energy level shifts in electronic hydrogen or from electron-proton scattering experiments. This 
so-called ''proton radius puzzle" has triggered a large activity and is  the subject of intense debate, see e.g. Refs.~\cite{Pohl:2013yb, Bernauer:2014cwa} for recent reviews. 
The limiting accuracy in extracting the proton charge radius from the Lamb shift measurements in muonic atoms is due to hadronic corrections. Among these, the leading uncertainty originates from the so-called polarizability correction, which corresponds with a TPE process between the lepton and the proton. This correction can be obtained from the knowledge of forward double virtual Compton structure amplitudes, which has been estimated from phenomenology~\cite{Pachucki:1996zza, Martynenko:2005rc, Carlson:2011zd, Gorchtein:2013yga}, 
from non-relativistic QED effective field theory~\cite{Hill:2011wy}, as well as from chiral effective field theory~\cite{Nevado:2007dd, Birse:2012eb, Alarcon:2013cba, Peset:2014jxa}.   
The total TPE corrections to the Lamb shift were found to be in the 10-15 \% range of the total discrepancy for the proton charge radius extractions between electron scattering and muonic atom spectroscopy. Although these TPE  corrections are not large enough to explain the bulk of the difference between both extraction methods, they constitute a large hadronic correction to the Lamb shift result, which needs to be taken into account as accurately as possible when extracting the proton radius from such experiments. 

The "proton radius puzzle" also calls for re-visiting the TPE corrections in the 
elastic electron-nucleon scattering data in the low-$ Q^2 $ region, from which  the proton radius is obtained. In the low $ Q^2 $ region we expect the main contribution to TPE corrections from the elastic intermediate state. Its leading contribution is given by the Coulomb scattering 
of relativistic electrons off the proton charge distribution, and was obtained by  McKinley and Feshbach \cite{McKinley:1948zz}.  Although the corrections to the Coulomb distortion in elastic electron-proton scattering were found to be small in the small $Q^2$ region~\cite{Blunden:2005jv}, a high precision extraction of the proton radii, especially its magnetic radius, calls for an assessment of the model dependence of the TPE corrections. 

In this work we aim to re-visit the TPE corrections in the region of low $Q^2$ up to about 1~GeV$^2$, and make a detailed comparison with the available data. In this work, 
we will focus our study on the elastic contribution of the TPE correction to the unpolarized elastic electron-proton scattering cross section. Two main calculations have been developed in the 
literature to estimate this elastic TPE contribution. A first method of calculation, performed by Blunden, Melnitchouk, and Tjon \cite{Blunden:2003sp} evaluates the two-photon box graph with the assumption of on-shell virtual photon-proton-proton vertices. A second method of calculation,  performed by Borisyuk and Kobushkin \cite{Borisyuk:2008es}, evaluates this elastic TPE 
correction within unsubtracted dispersion relations (DRs). In this work we compare these two methods and compare them in detail to the recent data.  
In order to minimize the model dependence due to unknown or poorly constrained contributions from higher intermediate states, we propose a DR approach with one subtraction, where the subtraction constant, which encodes the less well constrained physics at high energies, is fitted  to the available data. 

The paper is organized as follows: We describe the general formalism of elastic electron-proton scattering in the limit of massless electrons in Section~\ref{sec2}. We review the DR framework in Section~\ref{sec3}: we subsequently discuss how to obtain the imaginary parts of the TPE amplitudes from unitarity relations in the  physical region, their analytical continuation to the unphysical region, as well as how to reconstruct the real parts using dispersive integrals. 
We review the two-photon box graph model evaluation with the assumption of an on-shell form of virtual photon-proton-proton vertex in Section~\ref{sec7}: we subsequently discuss the loop diagram evaluation of the box graph, as well as its dispersive evaluation. We also discuss the 
forward limit and provide an analytical formula which describe the leading corrections beyond the Feshbach Coulomb correction formula.  In Section~\ref{sec8}, we make detailed comparisons 
between both methods, and show that a subtraction eliminates the differences between both methods. Using such a subtracted DR formalism for the TPE contribution, we provide a 
detailed study of available unpolarized and polarized elastic electron-proton scattering data for the case of the elastic intermediate state. 
We present our conclusions and outlook in Section~\ref{sec9}.
Some technical details on unitarity relations and on the integrals entering the box diagram are 
collected in three appendices. 
 
\section{Elastic electron-proton scattering in the limit of massless electrons}
\label{sec2}

\begin{figure}[h]
\begin{center}
\includegraphics[width=.4\textwidth]{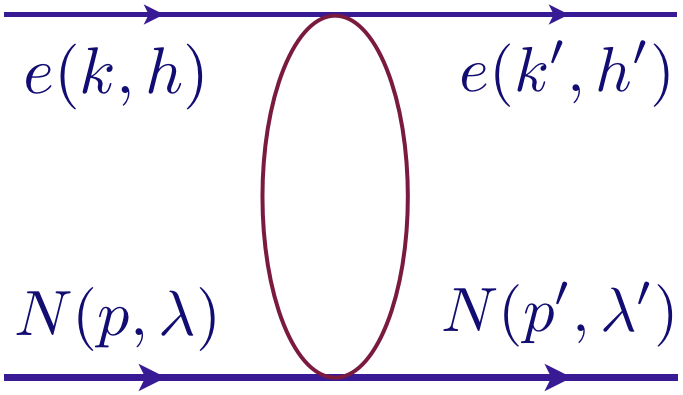}
\end{center}
\caption{Elastic electron-proton scattering.}
\label{ep_elastic}
\end{figure}

Elastic electron-proton scattering $ e( k , h ) + p( p, \lambda ) \to e( k', h') + p(p', \lambda') $,  where $ h(h') $ denote the incoming (outgoing) electron helicities and $ \lambda(\lambda') $ the corresponding proton helicities respectively, (see Fig. \ref{ep_elastic}) is completely described by 2 Mandelstam variables, e.g., $ Q^2 = - (k-k')^2 $ - the squared momentum transfer, and $ s = ( p + k )^2 $ - the squared energy in the electron-proton center-of-mass (c.m.) reference frame.

It is convenient to introduce the average momentum variables $ P = (p+p')/2, ~ K = (k+k')/2 $, the $ u $-channel squared energy $ u = ( k - p' )^2 $, and the crossing symmetry variable $ \nu = (s-u)/4 $ which changes sign under $ s \leftrightarrow u $ channel crossing. Instead of the Mandelstam invariant $ s $ or the crossing symmetric variable $ \nu $, it is customary in experiment to use the virtual photon polarization parameter $ \varepsilon $, which varies between 0 and 1, indicating the degree of the longitudinal polarization in case of one-photon exchange. We will be working in the limit of ultra-relativistic electrons, allowing to neglect the electron mass. In terms of $ Q^2 $ and $ \nu $, $ \varepsilon $ is then defined as
\ber
\label{epsilon}
\varepsilon & = & \frac{ 16 \nu^2 - Q^2 ( Q^2 + 4 M^2 )}{ 16 \nu^2 + Q^2 ( Q^2 + 4 M^2 )},
\eer
where $ M $ denotes the proton mass.

It is convenient to work in the c.m. reference frame with electron scattering angle $ \cma $. The momentum transfer is  then given by $ Q^2 = \frac{ (s - M^2)^2 }{s} \sin^2 \frac{\cma}{2} $.

There are 16 helicity amplitudes $ T_{h' \lambda', h \lambda} $ with arbitrary $h,h',\lambda,\lambda'$ = $\pm 1/2$ in Fig. \ref{ep_elastic}, but discrete symmetries of QCD and QED leave just six independent amplitudes. The momentum transfer accessed by current experiments down to $ Q^2 \gtrsim 0.001 ~ \mathrm{GeV}^2 $ \cite{Bernauer:2010wm, Bernauer:2013tpr} is still much larger than the squared electron mass, so that to very good approximation electrons can be treated as massless particles. As all amplitudes with electron  helicity flip are suppressed by the electron mass, in the limit of massless electrons only three independent helicity amplitudes survive: $ T_1 \equiv T_{\hp \hp, \hp \hp}, ~T_2 \equiv T_{\hp \hm, \hp \hp},  ~T_3 \equiv T_{\hp \hm, \hp \hm} $. 

The helicity amplitude for elastic $ e^{-} p $ scattering can be expressed through three independent tensor structures. It is common to use the following notations \cite{Guichon:2003qm}
\beq \label{FFs} 
T  =  \frac{e^2}{Q^2} \bar{u}(k',h) \gamma_\mu u(k,h) \cdot \bar{u}(p',\lambda') \left( \cG_M \gamma^\mu - \cF_2 \frac{P^{\mu}}{M} + \cF_3 \frac{\gamma . K P^{\mu}}{M^2} \right) u(p,\lambda),
\eeq
where the structure amplitudes $ \cG_M,~\cF_2,~\cF_3 $ are functions of $ \nu $ and $ Q^2 $.

Following the Jacob-Wick \cite{Jacob:1959at} phase convention, the three independent helicity amplitudes can be expressed through the structure amplitudes as
\ber \label{hel_ampl_st}
T_1 & = & 2 \frac{e^2}{Q^2} \left\{ \frac{ s u - M^4 }{s - M^2} \left(\cF_2 - \cG_M -  \frac{s-M^2}{2M^2} \cF_3 \right) + Q^2 \cG_M \right\}, \nonumber \\
T_2 & = &  - \frac{e^2}{Q^2}  \frac{\sqrt{Q^2 ( M^4 - s u )}}{M} \left\{ \cF_2 + 2 \frac{M^2}{s-M^2} \left(\cF_2 - \cG_M \right)-  \cF_3 \right\}  e^{ - i \phi}, \nonumber \\
T_3 & = & 2 \frac{e^2}{Q^2} \frac{ s u - M^4 }{s - M^2}  \left\{ \cF_2 - \cG_M - \frac{s-M^2}{2M^2} \cF_3 \right\},  
\eer
where $ \phi $ is the azimuthal angle of the scattered electron. Notice that following the Jacob-Wick phase convention, the azimuthal angular dependence of the helicity amplitudes is in general given by $ T_{h' \lambda', h \lambda} = e^{ i ( \Lambda - \Lambda' ) \phi }$, with $ \Lambda = h - \lambda $ and $ \Lambda' = h' - \lambda' $.    

The structure amplitudes can in turn be expressed through the helicity amplitudes as \cite{Pasquini:2004pv}
\ber \label{ff_s}
\cG_M & = & \frac{1}{2} \left\{ \tilde{t}_1 - \tilde{t}_3 \right\}, \nonumber \\
\cF_2 & = & \frac{ M Q }{\sqrt{ M^4 - s u}} \left\{ - \tilde{t}_2 e^{ i \phi} + \tilde{t}_3 \frac{ M Q }{\sqrt{ M^4 - s u}} \right\}, \nonumber \\
\cF_3 & = & \frac{M^2}{s - M^2} \left\{ - \tilde{t}_1 - \tilde{t}_2 \frac{ 2 M Q }{\sqrt{ M^4 - s u }} e^{ i \phi} + \tilde{t}_3 \left( 1 + Q^2\frac{ s + M^2 }{ M^4 - s u }\right) \right\}, 
\eer
with $ \tilde{t} = T/e^2 $.

In the one-photon ($ 1 \gamma$) exchange approximation the helicity amplitude for elastic $ e^{-} p $ scattering can be expressed in terms of the Dirac $ F_1 $ and Pauli $ F_2 $ FFs as
\beq \label{OPE_amplitude} 
T  =  \frac{e^2}{Q^2} \bar{u}(k',h) \gamma_\mu u(k,h) \cdot \bar{u}(p',\lambda') \left( \gamma^\mu F_1(Q^2) + \frac{i \sigma^{\mu \nu} q_\nu}{2 M} F_2(Q^2)  \right) u(p,\lambda).
\eeq
When extracting FFs from experiment, it is useful to introduce Sachs magnetic and electric FFs
\ber  \label{Sachs_ffs}
 G_M  =  F_1 + F_2 , ~~~~~~~
 G_E  = F_1 - \tau F_2 ,
\eer 
with $ \tau = Q^2 / (4 M^2) $.

In the one-photon exchange approximation, the structure amplitudes defined in Eq. (\ref{FFs}) can be expressed in terms of the FFs as: $ \cG_M = G_M (Q^2), ~\cF_2 = F_2 (Q^2), ~\cF_3 = 0 $. The exchange of more than one photon gives corrections to all amplitudes $ \cG_M,~\cF_2,~\cF_3 $, which we denote by 
\ber
\cG_M^{2 \gamma} &\equiv& \cG_M(\nu, Q^2) - G_M(Q^2), \nonumber \\
\cF_2^{2 \gamma} &\equiv& \cF_2(\nu, Q^2) - F_2(Q^2), \nonumber \\
\cF_3^{2 \gamma} &\equiv& \cF_3(\nu, Q^2). 
\eer 

In the following, we consider the correction to observables due to TPE which are corrections of order $ e^2 $. The correction to the unpolarized elastic electron-proton cross section is given by the interference between the $1 \gamma$-exchange diagram and the sum of box and crossed-box diagrams with two photons. Including the TPE corrections, we can express the $ e^{-} p $ elastic cross section through the cross section in the $1 \gamma$-exchange approximation $ \sigma_{1 \gamma} $ by 
\ber
 \sigma = \sigma_{1 \gamma} \left( 1 + \delta_{2 \gamma} \right), 
\eer
where the TPE correction $ \delta_{2 \gamma} $ can be expressed in terms of the TPE amplitudes as
\beq \label{delta_TPE}
\delta_{2 \gamma} = \frac{2}{G^2_M  + \frac{\varepsilon}{\tau} G^2_E } \left\{ \left( G_M + \frac{\varepsilon}{\tau} {G_E} \right) \Re  {\cal{G}}_{M}^{2 \gamma} - \frac{\varepsilon( 1 + \tau)}{\tau} {G_E}  { \Re  \cal{F}}_{2}^{2 \gamma}  + \left( G_M + \frac{1}{\tau} {G_E} \right) \frac{\nu \varepsilon}{M^2} {{\Re \cal{F}}_3^{2 \gamma}}  \right\} .
\eeq

The longitudinal and transverse polarization transfer observables ($ P_t $ and $ P_l $)  are also influenced by TPE. The following ratio is measured experimentally \cite{Meziane:2010xc}
\ber \label{polarization_observables1}
 - \sqrt{\frac{\tau ( 1 + \varepsilon )}{2 \varepsilon}} \frac{P_t}{P_l} & = & \frac{G_E}{G_M} + \left( 1 + \tau \right) \frac{ F_2 \Re \cG_M^{2 \gamma}  - G_M \Re \cF_2^{2 \gamma}  }{G_M^2} + \left( 1 - \frac{ 2 \varepsilon }{ 1 + \varepsilon } \frac{G_E}{G_M} \right) \frac{\nu}{M^2} \frac{ \Re \cF_3^{2 \gamma}  }{G_M}.
\eer

Experimental data on longitudinal polarization transfer allows to reconstruct \cite{Meziane:2010xc}
\ber \label{polarization_observables2}
 \frac{P_l}{P^{Born}_l} & = & 1 - \frac{2 \varepsilon}{1 + \frac{\varepsilon}{\tau} \frac{G^2_E}{G^2_M}}  \frac{ 1 + \tau }{\tau}  \frac{G_E}{G_M^3} \left( F_2 \Re \cG_M^{2 \gamma}  - G_M \Re \cF_2^{2 \gamma}  \right) \nonumber \\
 & - & \frac{2 \varepsilon}{1 + \frac{\varepsilon}{\tau} \frac{G^2_E}{G^2_M}} \left( \frac{ \varepsilon }{ 1 + \varepsilon } \left( 1 - \frac{G_E^2}{\tau G_M^2} \right) + \frac{G_E}{\tau G_M} \right) \frac{\nu}{M^2} \frac{ \Re \cF_3^{2 \gamma}  }{G_M}.
\eer

For further use, it will be convenient to introduce amplitudes $ \cG_1, \cG_2 $ defined as
\ber 
 {\cal{G}}_{1}^{2 \gamma}  & = & {\cal{G}}_{M}^{2 \gamma}  + \frac{\nu}{M^2} {\cal{F}}_3^{2 \gamma} ,  \label{new_amplitude1}  \\ 
 {\cal{G}}_{2}^{2 \gamma}  & = & {\cal{G}}_{M}^{2 \gamma}  - \left( 1 + \tau \right) {\cal{F}}_2^{2 \gamma}  + \frac{\nu}{M^2} {\cal{F}}_3^{2 \gamma} . \label{new_amplitude2}
\eer

In terms of these amplitudes, the TPE correction to the unpolarized cross section is given by
\beq \label{unpolarized_cross_section}
\delta_{2 \gamma} = \frac{2}{G^2_M  + \frac{\varepsilon}{\tau} G^2_E } \left\{ G_M  \Re  {\cal{G}}_{1}^{2 \gamma}  + \frac{\varepsilon}{\tau} {G_E}  { \Re  \cal{G}}_{2}^{2 \gamma}  +  {G_M} \left (\varepsilon -1\right) \frac{\nu}{M^2} {{\Re \cal{F}}_3^{2 \gamma}}  \right\}, 
\eeq
and the polarization transfer observables can be written as
\ber
\frac{P_t}{P_l} & = & - \sqrt{\frac{2 \varepsilon}{\tau ( 1 + \varepsilon )}} \left( \frac{G_E}{G_M} + \frac{ \Re \cG_2^{2 \gamma}  }{G_M} -\frac{G_E}{G_M} \frac{ \Re \cG_1^{2 \gamma}  }{G_M} + \frac{ 1 - \varepsilon }{ 1 + \varepsilon } \frac{G_E}{G_M} \frac{\nu}{M^2} \frac{ \Re \cF_3^{2 \gamma}  }{G_M} \right),  \label{polarization_observables1x} \\
 \frac{P_l}{P^{Born}_l} & = & 1 - \frac{2 \varepsilon}{1 + \frac{\varepsilon}{\tau} \frac{G^2_E}{G^2_M}} \left\{ \frac{G_E}{\tau G_M} \frac{ \Re \cG_2^{2 \gamma}  }{G_M} - \frac{G^2_E}{\tau G^2_M} \frac{ \Re \cG_1^{2 \gamma}  }{G_M} \right. \nonumber \\
 &&\hspace{2.5cm} \left. + \left( \frac{ \varepsilon }{ 1 + \varepsilon } + \frac{ 1 }{ 1 + \varepsilon } \frac{G^2_E}{\tau G^2_M} \right) \frac{\nu}{M^2} \frac{ \Re \cF_3^{2 \gamma}  }{G_M}\right\}. \label{polarization_observables2x}
\eer

\section{Dispersion relation formalism}
\label{sec3}

In this work, we will calculate the TPE corrections to the invariant amplitudes $ \cG_M^{2 \gamma} , \cF_2^{2 \gamma}  $ and $ \cF_3^{2 \gamma}  $ in a 
dispersion relation (DR) formalism. For simplicity of notation, we will drop the subscript 
${2 \gamma} $ on the invariant amplitudes in all of the following of this paper, and understand that we already subtracted off the $1\gamma$ parts. 

Assuming analyticity, one can write down DRs for the invariant amplitudes. 
As a consequence of Cauchy's theorem the real parts of the structure amplitudes can be obtained from the imaginary parts with the help of DRs expressed in the complex plane of the $ \nu $ variable for fixed value of momentum transfer $ Q^2 $. The imaginary parts of the amplitudes which enter the DRs are related using unitarity to physical observables. The DRs require the amplitudes to have a sufficiently falling behavior at high energies to ensure convergence, otherwise a subtraction is required.

In this section, we will set up the details of the DR formalism for the TPE contribution to elastic $ e^{-} p $ scattering, and apply it to the case of a proton intermediate state.

\subsection{Unitarity relation}
\label{sec4}
The imaginary parts of the invariant amplitudes can be obtained with the help of the unitarity equation for the scattering matrix $ S $ (with $ S = 1 + i T $)
\ber
 S^+ S  =  1 ,  ~~ T^+ T = i ( T^+ - T).
\eer

\begin{figure}[htp]
\begin{center}
\includegraphics[width=.85\textwidth]{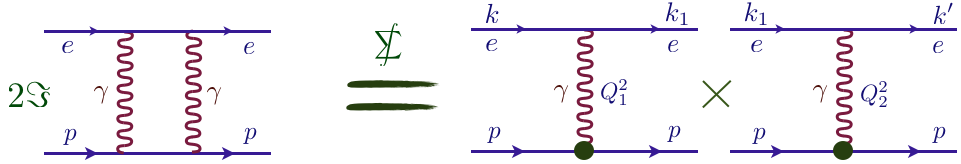}
\end{center}
\caption{Unitarity relations for the case of the elastic intermediate state contribution.}
\label{unitarity}
\end{figure}

For the numerical estimates in this paper, we will consider the unitarity relations for the nucleon intermediate state contribution, which by definition only involves on-shell amplitudes in the $1 \gamma$-exchange approximation. The unitarity relation is represented in Fig.~\ref{unitarity}.

In the c.m. frame, the electron energy is $ k_0 =  (s - M^2) / (2\sqrt{s}) $. The electron initial ($k$), intermediate ($k_1$) and final ($k'$) momentums are given by
\ber \label{kinematics_unitarity}
 k & = & k_0 (1,0, 0, 1) , \nonumber \\
 k_1 & = & k_0 (1, \sin \theta_1 \cos \phi_1 , \sin \theta_1 \sin \phi_1 , \cos \theta_1),\nonumber \\ 
 k' & = & k_0 (1, \sin \cma , 0, \cos \cma),  
\eer 
with intermediate electron angles $ \theta_1 $ and $ \phi_1 $.

We also introduce the relative angle between the 3-momentum of intermediate and final electrons as $ \hat{k}_1 \cdot \hat{k}' \equiv  \cos \theta_2 $, with $ \cos \theta_2 = \cos \cma \cos \theta_1 + \sin \cma \sin \theta_1 \cos \phi_1 $.

The imaginary parts of the $ 2 \gamma $-exchange helicity amplitudes are given by 
\ber \label{unitarity_im}
 \Im T_1 & = &\frac{1}{64 \pi^2} \frac{s-M^2}{s}  \mathop{\mathlarger{\int}} \left\{  T^{ 1 \gamma}_1(Q^2_1) T^{ 1 \gamma}_1(Q^2_2) +   T^{ 1 \gamma}_2(Q^2_1) T^{ 1 \gamma}_2(Q^2_2)  \cos(  \tilde{\phi}') \right\}  \mathrm{d} \Omega,\nonumber \\
 \Im T_3 & = &\frac{1}{64 \pi^2} \frac{s-M^2}{s}  \mathop{\mathlarger{\int}} \left\{  T^{ 1 \gamma}_3(Q^2_1) T^{ 1 \gamma}_3(Q^2_2)  \cos( \phi-\phi') -   T^{ 1 \gamma}_2(Q^2_1) T^{ 1 \gamma}_2(Q^2_2)  \cos( \phi + \tilde{\phi}) \right\}   \mathrm{d} \Omega, \nonumber   \\ 
 \Im T_2 & = & \frac{1}{64 \pi^2} \frac{s-M^2}{s}  \mathop{\mathlarger{\int}} \left\{ T^{ 1 \gamma}_2(Q^2_1) T^{ 1 \gamma}_3(Q^2_2)  \cos( \phi' )  + T^{ 1 \gamma}_1(Q^2_1) T^{ 1 \gamma}_2(Q^2_2)  \cos(  \tilde{\phi}) \right\}   \mathrm{d} \Omega,
\eer
where the phases $ \phi, \phi', \tilde{\phi}, \tilde{\phi}' $ are defined in Eq. (\ref{angles}) of Appendix \ref{app1}. The  momentum transfers $ Q^2_1 $ and $ Q^2_2 $ correspond with the scattering from initial to intermediate state and with the scattering from intermediate to final state respectively. The $1 \gamma$-exchange amplitudes, which were defined in Eq. (\ref{unitarity_im}) by explicitly taking out all kinematical phases, can be obtained from Eq. (\ref{hel_ampl_st}) after substitution of the structure amplitudes by the corresponding FFs: $ \cG_M \to G_M, ~\cF_2 \to F_2, ~\cF_3 \to 0$ and are given by
\ber \label{hel_ampl_st_OPE}
T^{1 \gamma}_1 & = & 2 \frac{e^2}{Q^2} \left\{ \frac{ s u - M^4 }{s - M^2} (F_2 - G_M) + Q^2 G_M \right\}, \nonumber \\
T^{1 \gamma}_2 & = &  - \frac{e^2}{Q^2}  \frac{\sqrt{Q^2 ( M^4 - s u )}}{M} \left\{ F_2 + 2 \frac{M^2}{s-M^2} (F_2 - G_M) \right\}, \nonumber \\
T^{1 \gamma}_3 & = & 2 \frac{e^2}{Q^2} \frac{ s u - M^4 }{s - M^2}  \left( F_2 - G_M \right).  
\eer

In case of the forward scattering $ \sin(\cma) = 0, ~ \cos(\cma) = 1, $ the unitarity relations lead to the optical theorem for amplitudes without helicity flip of the proton. The proton helicity-flip amplitude $ T_2 $ vanishes in this limit.

\subsection{Dispersion relations}
\label{sec5}

To discuss DRs for the invariant amplitudes describing the elastic $ e^{-} p $  scattering it is convenient to use amplitudes which have a definite behavior under $ s \leftrightarrow u $ crossing symmetry. In terms of the crossing symmetry variable $ \nu = ( s - u ) / 4 $, one can verify that the TPE invariant amplitudes have following crossing symmetry properties
\ber \label{crossing_relations}
& \cG_1(-\nu,Q^2) =  - \cG_1(\nu,Q^2) , ~~ \cG_2(-\nu,Q^2) =  - \cG_2(\nu,Q^2), \nonumber \\
& \cG_M(-\nu,Q^2) =  - \cG_M(\nu,Q^2) , ~~ \cF_{2}(-\nu,Q^2) = - \cF_{2}(\nu,Q^2), \nonumber \\
& \cF_{3}(-\nu,Q^2) = \cF_{3}(\nu,Q^2). 
\eer

\begin{figure}[h]
\begin{center}
\includegraphics[width=.55\textwidth]{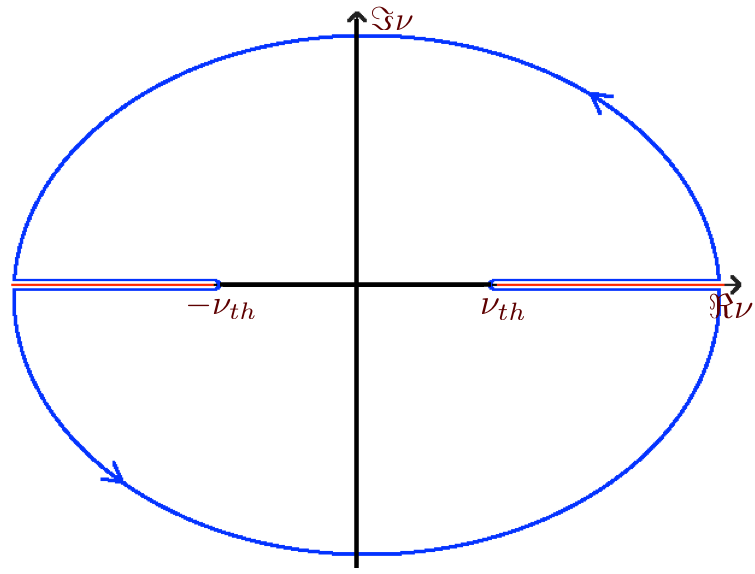}
\end{center}
\caption{Complex plane of the $ \nu $ variable.}
\label{complex_plane}
\end{figure}

 The general form of fixed-$Q^2$ DR for the function with definite crossing symmetry properties can be obtained from the complex plane shown in Fig. \ref{complex_plane} and is given by
\ber 
  \Re \cG(\nu, Q^2) = \frac{1}{\pi} \left( \mathop{\mathlarger{\int}} \limits^{~~ \infty}_{\nu_{th}} \frac{\Im \cG (\nu'+ i 0, Q^2)}{\nu'-\nu}  \mathrm{d} \nu' - \mathop{\mathlarger{\int}} \limits^{~~ -\nu_{th}}_{-\infty} \frac{\Im \cG (\nu'- i 0, Q^2)}{\nu'-\nu}  \mathrm{d} \nu' \right).
 \eer

The dispersive integral starts from the threshold $\nu_{th} $ corresponding with the cut. The threshold corresponding with the elastic cut due to the nucleon intermediate state is located at $ s = M^2 $ or  $ \nu_{th} = \nu_B = -Q^2/4 $, so there is an integration region with intersection of  $ s $- and $ u $-channel cuts. The threshold corresponding with the inelastic cut due to the pion-nucleon intermediate states is given by: $ s = (M+m_\pi)^2 $ or $ \nu_{th} = m_{\pi} ( m_{\pi} + 2 M )/2 -Q^2/4 $.

The amplitudes which are odd in $ \nu $, $ {\cal{G}}^{odd} $, satisfy
\ber
 \label{oddDR}
 \Re {\cal{G}}^{odd}(\nu, Q^2) & = & \frac{2}{\pi} \nu  \mathop{\mathlarger{\int}} \limits^{~~ \infty}_{\nu_{th}} \frac{\Im {\cal{G}}^{odd} (\nu'+ i 0, Q^2)}{\nu'^2-\nu^2}  \mathrm{d} \nu' .
 \eer
 The amplitudes which are even in $ \nu$, $ {\cal{G}}^{even} $, satisfy
 \ber
 \label{evenDR}
 \Re  {\cal{G}}^{even} (\nu, Q^2) & = & \frac{2}{\pi} \mathop{\mathlarger{\int}} \limits^{~~ \infty}_{\nu_{th}}  \nu' \frac{\Im  {\cal{G}}^{even}  (\nu'+ i 0, Q^2)}{\nu'^2-\nu^2}  \mathrm{d} \nu' .
 \eer
Unsubtracted DRs as given by Eqs. (\ref{oddDR}, \ref{evenDR})  can only be written down for functions with appropriate high-energy (HE) behavior. We will next discuss the HE behavior of the structure amplitudes for the case of the box diagram calculation with nucleon intermediate state, which will be explained in detail in Section \ref{sec7}.

For the discussions of the HE behavior in the box diagram model with nucleon intermediate state, we consider the virtual photon-proton-proton vertices as point couplings. Furthemore, we consider three contributions, whether both vertices correspond with vector couplings (referred to as $ \mathrm{F}_1 \mathrm{F}_1 $ structure), both vertices correspond with tensor couplings ($ \mathrm{F}_2 \mathrm{F}_2 $ structure), or whether one vertex corresponds with a vector and the second vertex with a tensor coupling ($ \mathrm{F}_1 \mathrm{F}_2 $ structure). 

In general, the HE behavior ($ \nu >> Q^2, ~M^2 $) of the amplitudes can be parametrized as $ \cG (\nu)  \simeq  \left( c_1 \nu^{x_1}  + c_2 \nu^{x_2} \ln \nu \right) $, where the parameters can be extracted from a fit to the calculation. In Tables l-lll, we show the extracted values of the powers $x_1, ~x_2$ for the different structure amplitudes and for the different cases of virtual photon-proton-proton vertices.

\begin{table}[h] \label{he_F1F1}
\begin{tabular}{|r|r|r|r|r|r||r|r|r|r|r|r|}
\hline
& $ \Im \cG_M $ & $ \Im \cF_2 $ & $ \Im \cF_3 $ & $ \Im \cG_1 $ & $ \Im \cG_2 $ & $ \Re \cG_M $ & $ \Re \cF_2 $ & $ \Re \cF_3 $ & $ \Re \cG_1 $ & $ \Re \cG_2 $ \\ \hline
$x_1$ & 0 & -2 & -1 & -1 & -1 & 0 & -1 & -1 & -1 & -1 \\ \hline
$x_2 $& 0 & -2 & -1 & -1 & -1 & 0 & -1 & -1 & -1 & -1 \\ \hline
\end{tabular}
\caption{The values of the powers $x_1$ and $x_2$ in the HE fit of the different structure amplitudes according to the form $ \cG (\nu)  \simeq  \left( c_1 \nu^{x_1}  + c_2 \nu^{x_2} \ln \nu \right) $, for the box diagram model with point-like $ \mathrm{F}_1 \mathrm{F}_1 $ vertex structure.}
\end{table}

\begin{table}[h] \label{he_F1F2}
\begin{tabular}{|r|r|r|r|r|r||r|r|r|r|r|r|}
\hline
& $ \Im \cG_M $ & $ \Im \cF_2 $ & $ \Im \cF_3 $ & $ \Im \cG_1 $ & $ \Im \cG_2 $ & $ \Re \cG_M $ & $ \Re \cF_2 $ & $ \Re \cF_3 $ & $ \Re \cG_1 $ & $ \Re \cG_2 $ \\ \hline
$ x_1 $ & 0 & -1 & -1 & -1 & -1 & 0 & -1 & -1 & -1 & -1 \\ \hline
$ x_2$ & 0 & -1 & -1 & -1 & -1 & 0 & -1 & -1 & -1 & -1 \\ \hline
\end{tabular}
\caption{Same as Tab. l, but for the box diagram model with point-like $ \mathrm{F}_1 \mathrm{F}_2 $ vertex structure.}
\end{table}

\begin{table}[h] \label{he_F2F2}
\begin{tabular}{|r|r|r|r|r|r||r|r|r|r|r|r|}
\hline
& $ \Im \cG_M $ & $ \Im \cF_2 $ & $ \Im \cF_3 $ & $ \Im \cG_1 $ & $ \Im \cG_2 $ & $ \Re \cG_M $ & $ \Re \cF_2 $ & $ \Re \cF_3 $ & $ \Re \cG_1 $ & $ \Re \cG_2 $ \\ \hline
$x_1$ &  0 & -1 & -1 & 0 & 0 & 1 & -1 & 0 & 0 & 0 \\ \hline
$x_2$ &  0 & -1 & -1 & 0 & 0 & 0 & -1 & -1 & -1 & -1 \\ \hline
\end{tabular}
\caption{Same as Tab. l, but for the box diagram model with point-like $ \mathrm{F}_2 \mathrm{F}_2 $ vertex structure.}
\end{table}

For the case of $ \mathrm{F}_1 \mathrm{F}_1 $ and $ \mathrm{F}_1 \mathrm{F}_2 $ vertex structures, one notices that the behaviors of all amplitudes are sufficient to ensure unsubtracted DRs. For the case of two magnetic vertices ($ \mathrm{F}_2 \mathrm{F}_2 $ structure), we notice that the $ \cF_2, ~\cG_1, ~\cG_2 $ amplitudes are sufficiently convergent to satisfy an unsubtracted DR. However, after UV regularization the amplitude $ \cG_M $ ($\cF_3$) has a real part which is behaving as $\nu$ ($\nu^0$) respectively, which in both cases leads to a constant contribution due to the contour at infinity in Cauchy's integral formula. This constant term cannot be reconstructed from the imaginary part of the amplitude. To avoid such unknown contribution, we will use in our following calculations instead of the amplitudes $ \cG_M $ and $ \cF_2 $, which are odd in $ \nu $, the amplitudes $ \cG_1 $ and $ \cG_2 $, defined in Eqs. (\ref{new_amplitude1}, \ref{new_amplitude2}). As is clear from the Tables I-III, the amplitudes $ \cG_1 $ and $ \cG_2 $ both satisfy unsubtracted DRs.
 
 For the amplitude $ \cF_3 $, which is even in $ \nu$, and for which an UV regularization has to be performed in the box diagram model when using point-like couplings, we will in the following compare the unsubtracted DR with a once-subtracted DR, with subtraction at a low energy point $ \nu_0 $, of the form~:
\ber
 \label{subDR}
 \Re {\cal{G}}^{even}(\nu, Q^2) - \Re {\cal{G}}^{even}(\nu_0, Q^2) & = & \frac{2 \left( \nu^2 - \nu_0^2 \right)}{\pi} \mathop{\mathlarger{\int}} \limits^{~~ \infty}_{\nu_{th}} \frac{ \nu'  \Im {\cal{G}}^{even} (\nu'+ i 0, Q^2)}{\left( \nu'^2-\nu^2 \right)\left(\nu'^2-\nu_0^2 \right )}  \mathrm{d} \nu'.   
 \eer

\subsection{Analytical continuation into the unphysical region}
\label{sec6}

\begin{figure}[htp]
\begin{center}
\includegraphics[width=.9\textwidth]{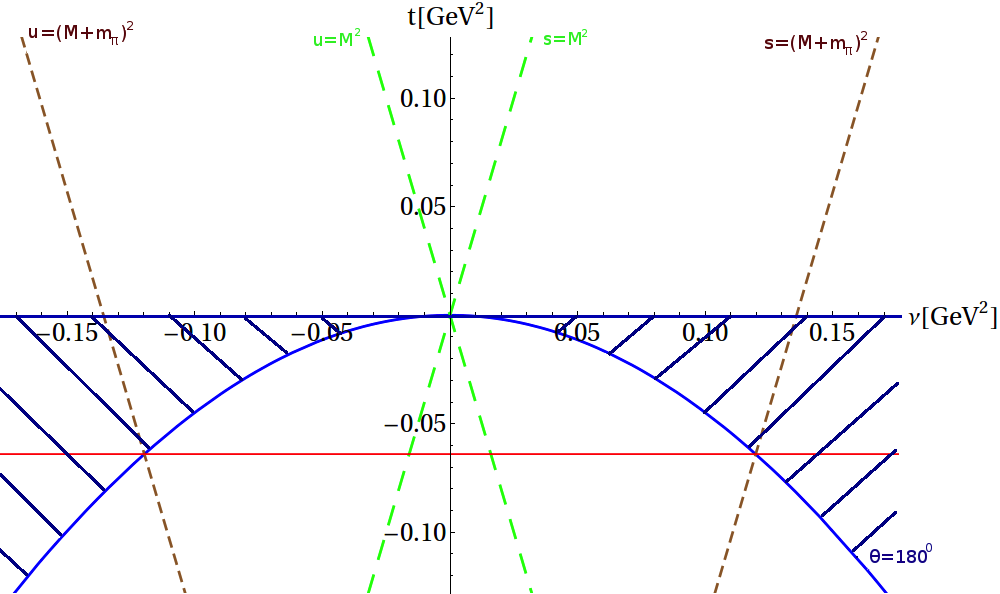}
\end{center}
\caption{Physical and unphysical regions of the kinematical variables $ \nu $ and $ t = - Q^2 $ (Mandelstam plot). The hatched blue region corresponds to the physical region, the long-dashed green lines give the elastic threshold positions, the short-dashed brown lines give the inelastic threshold positions. The horizontal red curve indicates the line at fixed negative $ t $ along which the dispersive integrals are evaluated. }
\label{mandelstam}
\end{figure}

To evaluate the dispersive integral at a fixed value of momentum transfer $ t = - Q^2 $ we have to know the imaginary part of the structure amplitude from the threshold in energy onwards. The imaginary part evaluated from the unitarity relations by performing a phase space integration over physical angles covers only the "physical" region of integration. The structure amplitudes also have an imaginary part outside the physical region as long as one is above the threshold energy. Accounting for only the contribution of the physical region to the structure amplitudes is in contradiction with the results obtained from the direct box graph evaluation for the electron-muon scattering \cite{VanNieuwenhuizen:1971yn}. Starting from the imaginary part of the structure amplitude in the physical region, we will now discuss how to continue it analytically into the unphysical region. To illustrate the physical and unphysical regions, we show in Fig. \ref{mandelstam} the Mandelstam plot for elastic electron-proton scattering in the limit of massless electrons. 
 
The threshold of the physical region is defined by the hyperbola
\ber \label{nuphys}
\nu = \nu_{ph} \equiv \frac{\sqrt{Q^2(Q^2+4M^2)}}{4}. 
\eer
Therefore, the evaluation of the dispersive integral for the elastic intermediate state at $ t = - Q^2 < 0 $ always requires information from the unphysical region. Note that for $ -t = Q^2 < 4 m^2_\pi (1 + \frac{m_\pi}{2M})^2/(1 + \frac{m_\pi}{M})^2 \simeq 0.064 ~ \mathrm{GeV}^2 $ (indicated by the red horizontal line in Fig. \ref{mandelstam}) an analytical continuation into the unphysical region is only required for the evaluation of the cut in the box diagram due to the nucleon intermediate states. For $ Q^2 $ larger than this value, also the evaluation of the cut due to the $ \pi N $ inelastic intermediate states requires an analytical continuation into the unphysical region.

\begin{figure}[htp]
\begin{center}
\includegraphics[width=.65\textwidth]{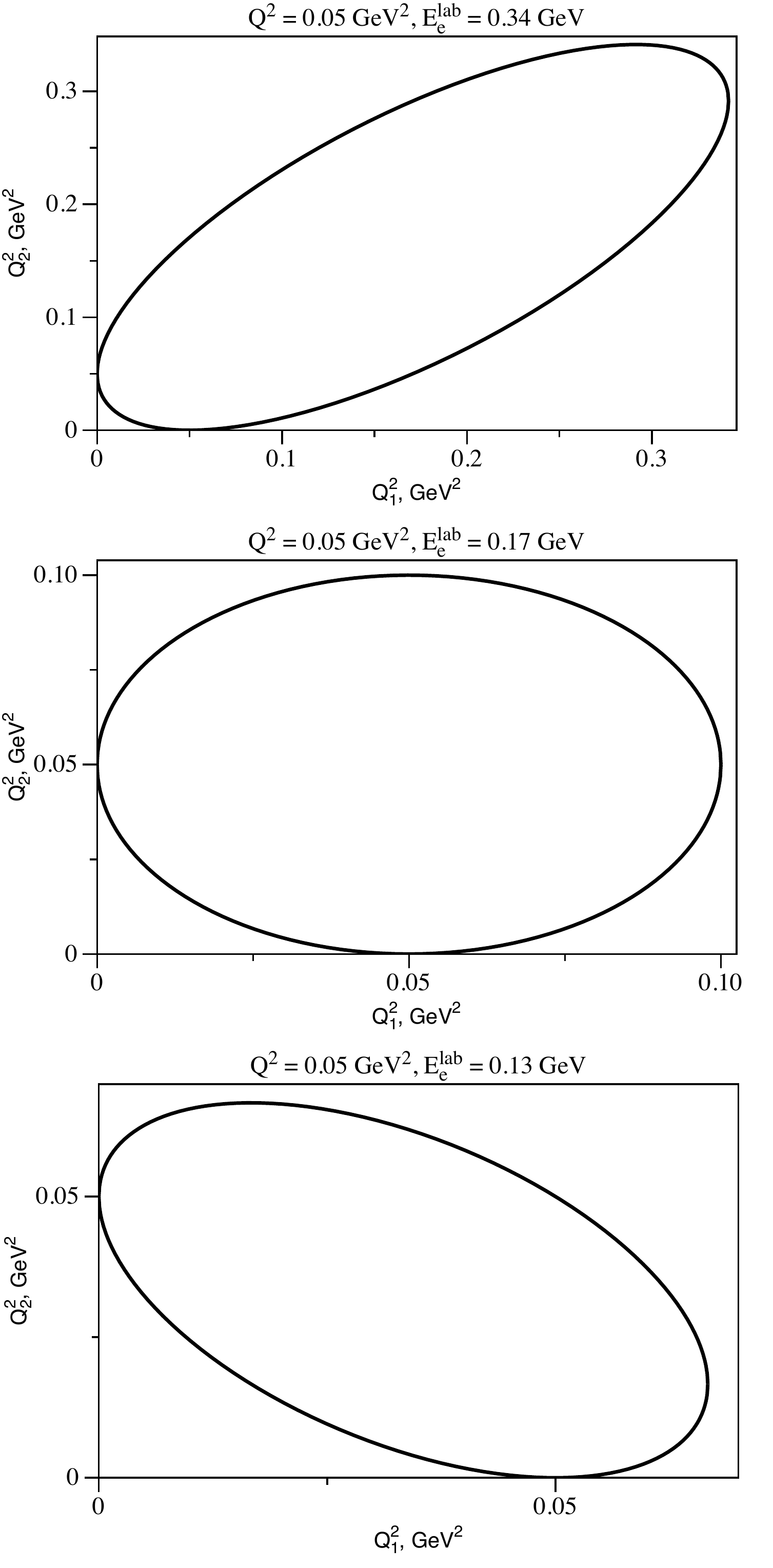}
\end{center}
\caption{The phase space integration regions entering the unitarity relations for the case of a nucleon intermediate state.}
\label{regions}
\end{figure}

We next discuss the integration region entering the unitarity relations for the case of the nucleon intermediate state contribution. The momentum transfers for the $ 1 \gamma$-exchange processes entering the r.h.s. of the unitarity relations Eq. (\ref{unitarity_im}) are given by
\beq
 Q^2_1 =  \frac{(s-M^2)^2}{2s} \left( 1 - \cos \theta_1 \right), ~~~ Q^2_2 =  \frac{(s-M^2)^2}{2s} \left( 1 - \cos \theta_2 \right).
\eeq
The indices 1, 2 correspond to scattering from initial to intermediate state and from intermediate to final state. The momentum transfer $ Q^2 $ obtains its maximal value for backward scattering $ \theta = 180^0 $. If $ Q^2_i $ is maximal (i.e., $ \theta_i = 180^{0} $), then $ Q^2_2 $ can be evaluated as
\ber
Q^2_{1} & = & Q^2_{ max} = \frac{\left(s-M^2\right)^2}{s}, \nonumber \\
Q^2_{2} & = & \frac{1}{s} \left( \left(s-M^2\right)^2 - s Q^2 \right).
\eer
The phase space integration in Eq. (\ref{unitarity_im}) maps out an ellipse in the $ Q_1^2, ~Q^2_2 $ plane, where the position of the major axis depends on the elastic scattering angle (or $Q^2$). The centre of the ellipse is located at $ Q^2_1 = Q^2_2 = Q^2_{max}/2 \equiv Q^2_c $. For forward and backward scattering, the ellipse reduces to a line: $ Q^2_1 = Q^2_2 $ for $ \cma= 0^0 $, and $ Q^2_2 = Q^2_{max} - Q^2_1 $ for $ \cma = 180^0 $. In Fig. \ref{regions}, we show the physical integration regions for different elastic scattering kinematics which we will consider in this work (the electron energy in the lab frame $ E^{lab}_e $, corresponding with a fixed target, is related to the $ s $ variable by $ s = M^2 + 2 M E^{lab}_e $).

\begin{figure}[htp]
\begin{center}
\includegraphics[width=.65\textwidth]{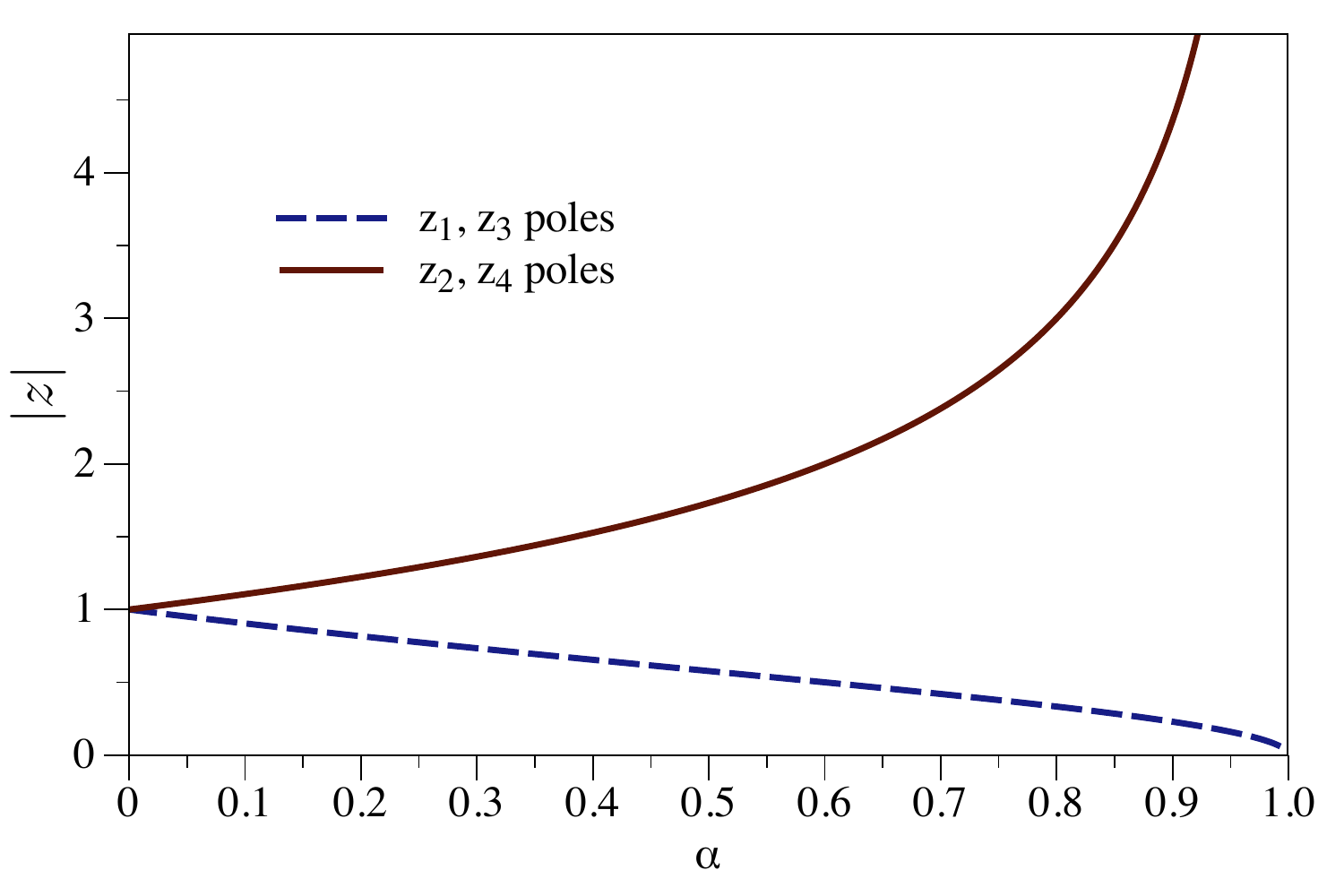}
\end{center}
\caption{The moduli of the pole positions in the physical region entering the angular integral in Eq. (\ref{ph_space}) for $ E_e^{lab} = 0.3 ~ \mathrm{GeV} $, $ \mu = 10^{-6} ~ \mathrm{GeV} $. Note that these moduli do not depend on the momentum transfer $ Q^2 $. The poles $z_1$ and $ z_3$ are inside the unit circle of integration ($ |z| = 1 $) for all values of $ \alpha $.}
\label{poles_physical_region}
\end{figure}

We will now demonstrate the procedure of analytical continuation on the example of the integral which corresponds with one denominator (originating from one of both photon propagators) on the r.h.s. of the unitarity relations Eq. (\ref{unitarity_im}). We introduce a small photon mass $ \mu $ to regulate IR singularities. The phase space integration entering the unitarity relations can be expressed in terms of elliptic coordinates $ \alpha $ and $ \phi $ (see Appendix \ref{app2}) as
\ber
 \mathop{\mathlarger{\int}} \frac{g(Q^2_1,Q^2_2)  \mathrm{d} \Omega }{Q^2_{1,2}+\mu^2}  & \sim &  \mathop{\mathlarger{\int}} \limits^{~~1}_{0} \mathrm{d} \alpha \mathop{\mathlarger{\int}}  \limits^{~~2\pi}_0 \mathrm{d} \phi \frac{g\left(Q^2_c ( a + b \cos \phi - c \sin \phi),Q^2_c ( a + b \cos \phi + c \sin \phi)\right)}{a + b \cos \phi \mp c \sin \phi},
\eer
with
\ber
 a = 1 + \frac{2 s \mu^2}{(s-M^2)^2}, ~~~b = \sqrt{1-\alpha^2} \sqrt{1- \frac{s Q^2}{(s-M^2)^2}}, ~~~ c =  \sqrt{1-\alpha^2} \sqrt{ \frac{s Q^2}{(s-M^2)^2}}. \nonumber
\eer
The angular integration can be performed on a unit circle in a complex plane with $ z = e^{ i \phi } $
\ber \label{ph_space}
 \mathop{\mathlarger{\int}} \limits^{~~2\pi}_0  \mathrm{d} \phi \frac{g(Q^2_1,Q^2_2)}{a + b \cos \phi - c \sin \phi} & = & -i \oint \frac{g(Q^2_1,Q^2_2)}{b + i c}  \frac{ 2 \mathrm{d} z}{(z - z_1)( z - z_2)}, \nonumber \\
 \mathop{\mathlarger{\int}} \limits^{~~2\pi}_0  \mathrm{d} \phi \frac{g(Q^2_1,Q^2_2)}{a + b \cos \phi + c \sin \phi} &  =  & -i \oint \frac{g(Q^2_1,Q^2_2)}{b - i c}  \frac{ 2 \mathrm{d} z}{(z - z_3)( z - z_4)},
\eer
with poles position given by 
\ber \label{poles1}
 z_{1,2} & = & \frac{1}{b + ic} ( -a \pm \sqrt{a^2 - (1-\alpha^2)}), 
\eer
\ber \label{poles2}
 z_{3,4} & = & \frac{1}{b - ic} (  -a \pm \sqrt{a^2 - (1-\alpha^2)}).
\eer
In the physical region $ ( s - M^2 )^2 > s Q^2 $, the integral is given by the residues of the poles $ z_1, z_3 $ ("+" sign in Eqs. (\ref{poles1}, \ref{poles2})), see Fig. \ref{poles_physical_region}.

\begin{figure}[htp]
\begin{center}
\includegraphics[width=.65\textwidth]{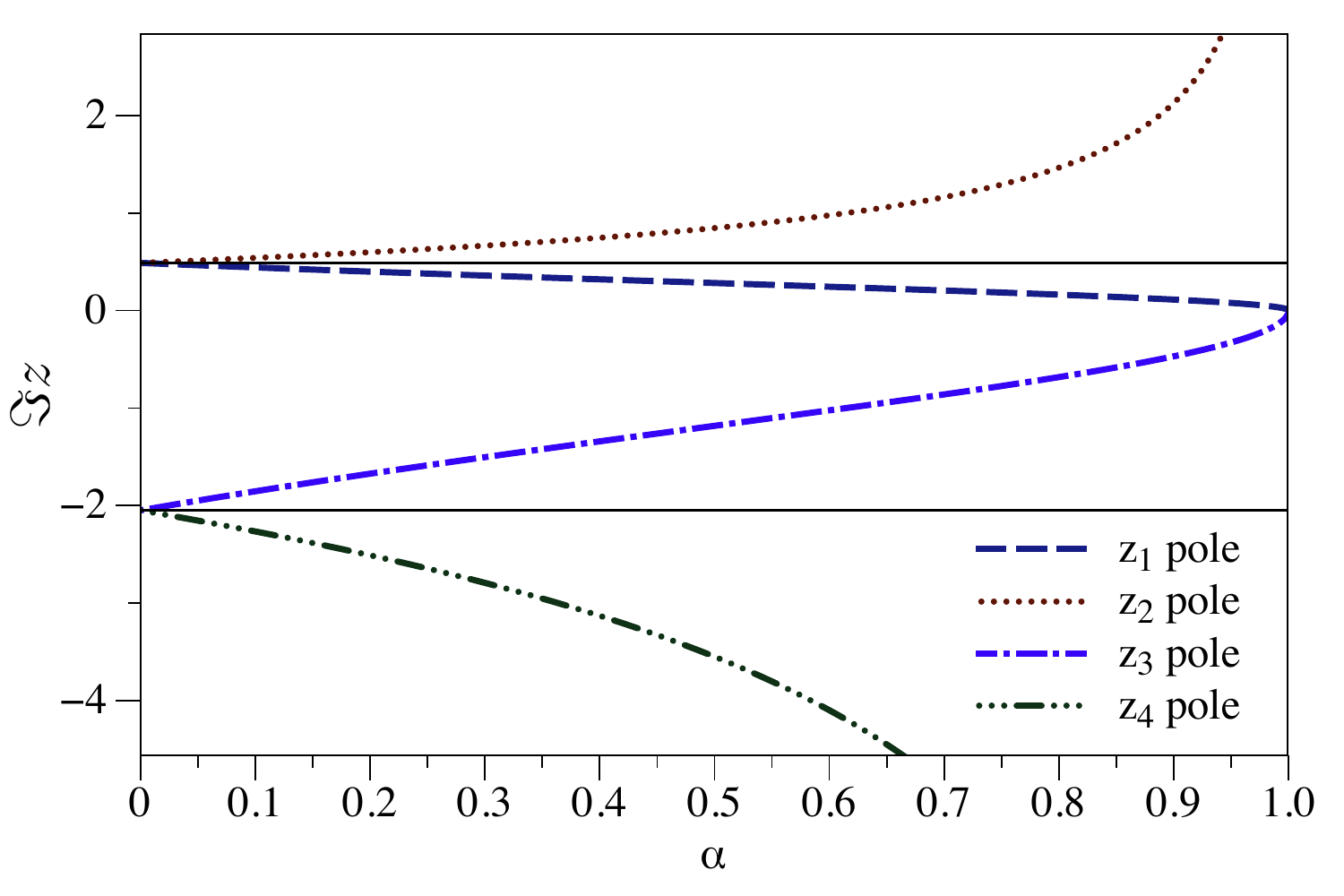}
\end{center}
\caption{ Imaginary part of the poles in the unphysical region entering the angular integral in Eq. (\ref{ph_space}) for $ E_e^{lab} = 0.3 ~ \mathrm{GeV} $, $ \mu = 10^{-6} ~ \mathrm{GeV} $ , $ Q^2 = 0.35 ~ \mathrm{GeV}^2$ (for which $ b_0 = 0.78 $ and $ c_0 = 1.27 $). The poles lie on the imaginary axis in the unphysical region. The pole $ z_3 $ is outside the unit circle for the values $ \alpha < \alpha_0 = 0.61 $. The intersections of the new contour of integration with the imaginary axis are shown by the horizontal solid lines, corresponding with values $ c_0 - b_0 \simeq 0.49$ (upper line) and $ - c_0 - b_0 \simeq 2.05$ (lower line) respectively.}
\label{poles_unphysical_region}
\end{figure}

In the unphysical region $ ( s - M^2 )^2 < s Q^2 $, the positions of the poles change with respect to the unit circle (Fig. \ref{poles_unphysical_region}), so the integral has a discontinuity at the transition point. To avoid the discontinuities, we define an analytical continuation by deforming the integration contour so as to include the poles $ z_1 $ and $ z_3 $. The integration can be done on the circle of the radius $ c_0 $ and the centre $ - i b_0 $ as
\ber
 \mathop{\mathlarger{\int}} \limits^{~~2 \pi}_0 f(e^{i\phi})  \mathrm{d} \phi & = & \mathop{\mathlarger{\oint}} \limits_{|z|=1} - i f(z) \frac{  \mathrm{d} z}{z} \to \mathop{\mathlarger{\oint}} \limits_{z=c_0 e^{i \phi} - i b_0} - i f(z) \frac{  \mathrm{d} z}{z},
\eer
with 
\beq
 c_0 = \sqrt{ s Q^2/(s-M^2)^2}, ~~~ b_0 = \sqrt{-1+ s Q^2/(s-M^2)^2}. \nonumber
\eeq

\begin{figure}[htp]
\begin{center}
\includegraphics[width=.65\textwidth]{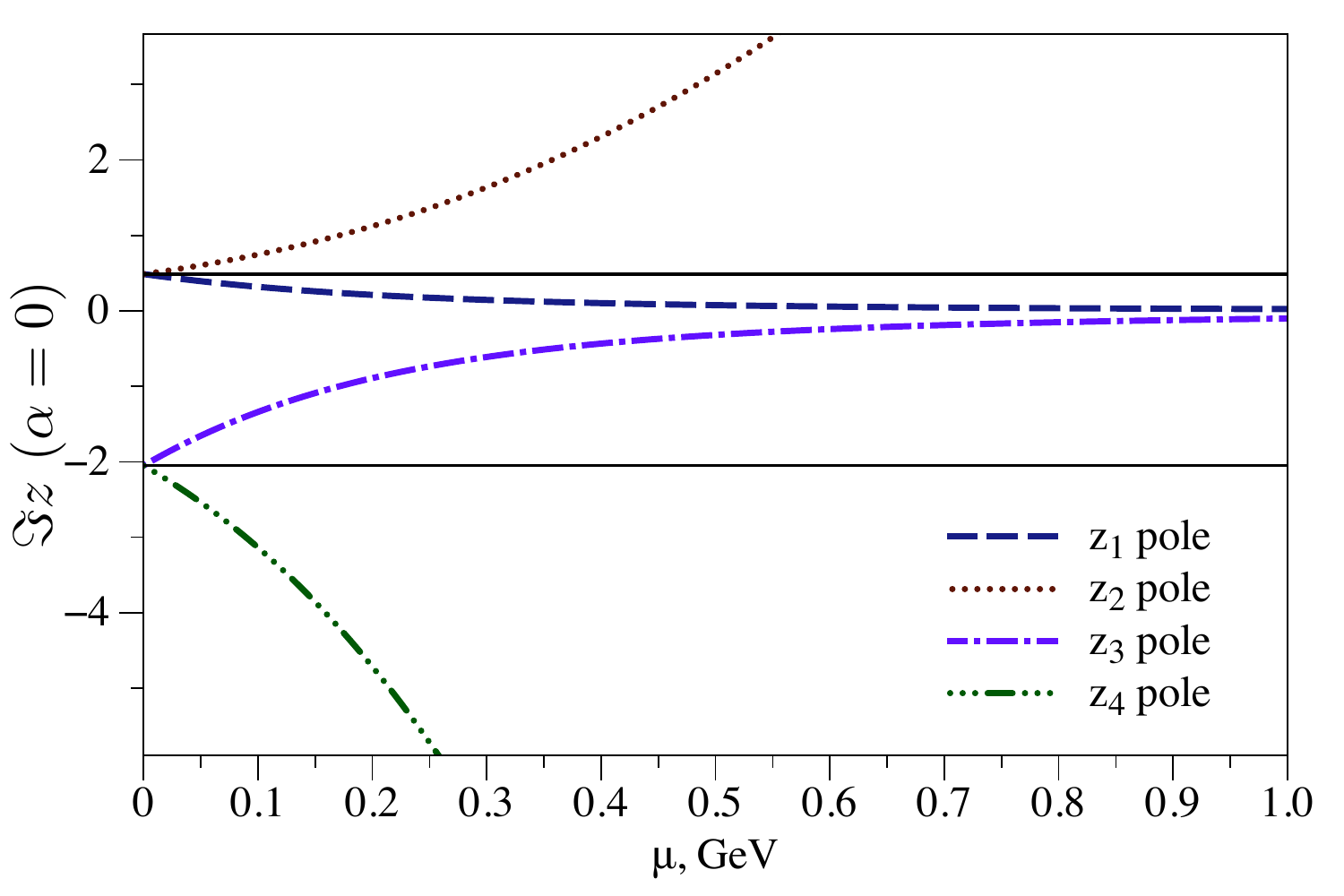}
\end{center}
\caption{ Same as Fig. \ref{poles_unphysical_region} for $ \alpha = 0 $ as function of $ \mu$.}
\label{poles_mu_dependence}
\end{figure}

For the value $ \alpha = 0 $, when the expression in brackets of Eqs. (\ref{poles1}, \ref{poles2}) approaches its minimum, the positions of the poles of interest (for small photon mass parameter $ \mu \to 0 $) are given by 
\ber
z_1 & = & \frac{i}{b_0 + c_0} \left( 1 - \frac{2 \mu \sqrt{s}}{s-M^2} \right) = i \left( c_0 - b_0 \right) \left( 1 - \frac{2 \mu \sqrt{s}}{s-M^2} \right), \nonumber \\
z_3 & = & \frac{i}{b_0 - c_0} \left( 1 - \frac{2 \mu \sqrt{s}}{s-M^2} \right) = - i \left( c_0 + b_0 \right) \left( 1 - \frac{2 \mu \sqrt{s}}{s-M^2} \right) .
\eer
These poles lie inside the deformed contour of integration which intersects the imaginary axis at $ \Im z = c_0 - b_0 $ and $ \Im z = - c_0 - b_0 $ respectively. We show in Fig. \ref{poles_mu_dependence} that with the growth of photon mass parameter $ \mu $ the poles move further away from the boundary of the integration region and therefore lie inside the new contour of integration.

\begin{figure}[htp]
\begin{center}
\includegraphics[width=.65\textwidth]{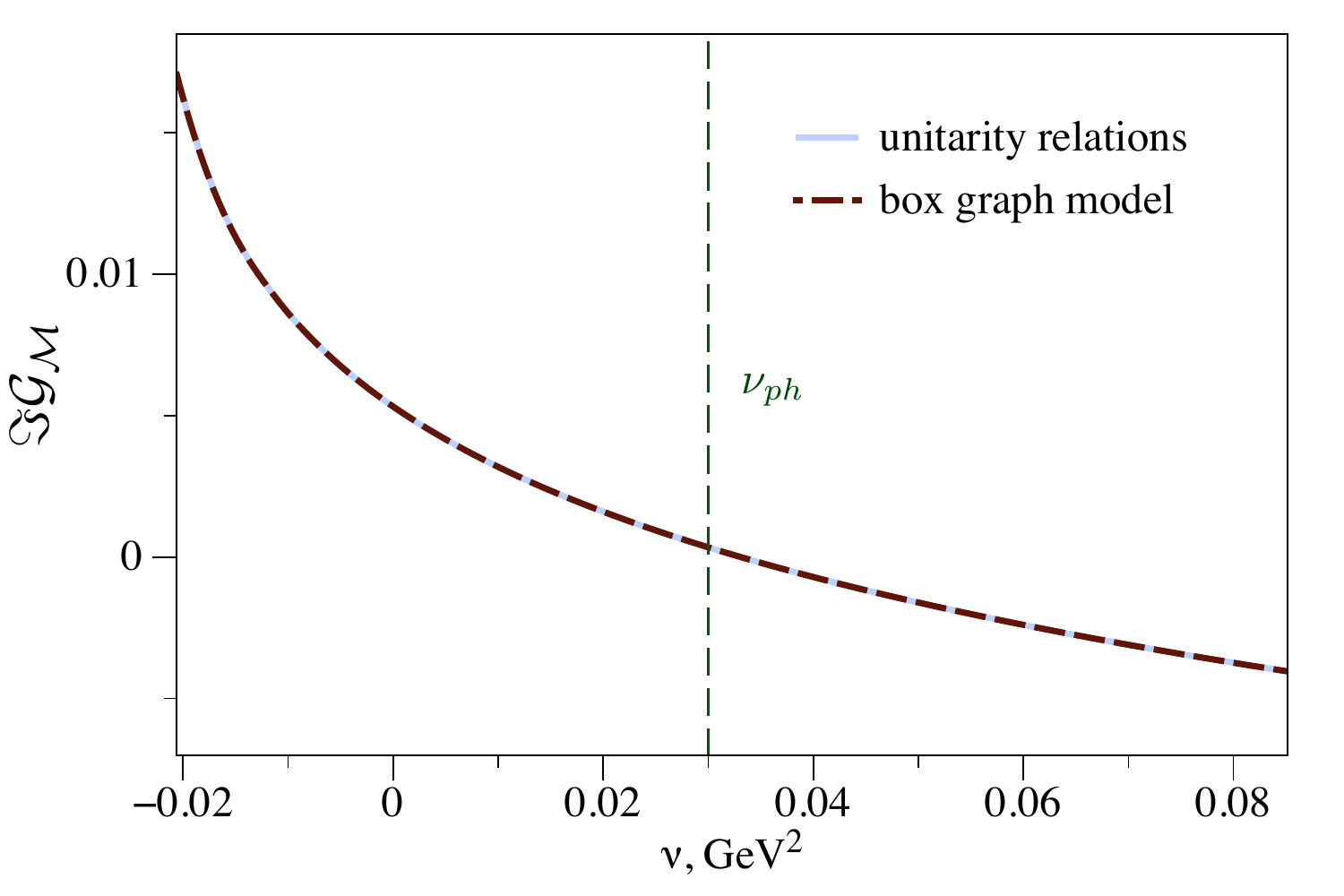}
\end{center}
\caption{Comparison between two evaluations of the imaginary part of the structure amplitude $ \cal{G}_M $ for $ e^{-} \mu^{-} $ scattering for $ Q^2 = 0.1 ~\mathrm{GeV}^2 $ corresponding with $ \nu_{ph} = 0.03 ~\mathrm{GeV}^2 $. Dashed-dotted curve: box graph evaluation; solid curve (coinciding): evaluation based on the unitarity relations. The region $ \nu > \nu_{ph} $ ($ \nu < \nu_{ph} $) corresponds with the physical (unphysical) region respectively.}
\label{imaginary_e_mu_plot}
\end{figure}

The deformed contour includes poles from both photon propagators, consequently the procedure of analytical continuation works also for two photon propagators in the unitarity relations Eq. (\ref{unitarity_im}). Therefore, through analytical continuation, the unitarity relations are able to reproduce the imaginary part of the structure amplitudes in the unphysical region also. As a cross-check of our procedure, we show the imaginary part $ \cG_M $ for the case of electron-muon scattering in Fig. \ref{imaginary_e_mu_plot}, as calculated using the analytically continued phase space integral, and compare it with the direct loop graph evaluation as explained in Section \ref{sec7} \cite{VanNieuwenhuizen:1971yn}. We find a perfect agreement between both calculations, justifying our analytical continuation procedure for the calculation based on unitarity relations.

A more realistic description of the proton is obtained by including electromagnetic FFs of the dipole form. This induces additional poles for the time-like region $ Q_i^2 < 0 $ in the unitarity relations Eq. (\ref{unitarity_im})
\ber
 G_M & \sim & \frac{1}{(Q_i^2 + \Lambda^2)^2}, ~~~~~ F_2 \sim \frac{1}{(Q_i^2 + 4 M^2)(Q_i^2 + \Lambda^2)^2}.
\eer
These poles arise from the dipole mass parameter $ \Lambda$ ($ Q_i^2 + \Lambda^2 = 0 $) and from the "kinematic" pole ($ Q_i^2 + 4 M^2 = 0 $). These poles can be treated in a similar way as the poles in Eqs. (\ref{poles1}, \ref{poles2}) through the replacement $ \mu \to \Lambda $ or $ \mu \to 2M $. These poles lie on the same line in the complex $ z $ plane as the $ z_1, z_2, z_3, z_4 $ poles. As soon as $ \Lambda > \mu, ~2M > \mu $, the new poles satisfy $ |z_1'| < |z_1|, ~|z_3'| < |z_3|,~ |z_2'| > |z_2|,~ |z_4'| > |z_4|$. From Fig. \ref{poles_mu_dependence}, where the $ \mu $ dependence of the pole positions in the unphysical region is shown, we see that our procedure of analytical continuation does not change the position of the new poles with respect to the deformed integration contour after the transition to the unphysical region. We can therefore conclude that the outlined procedure of analytical continuation is also valid for the calculation with proton FFs.



\section{Box diagram model calculation}
\label{sec7}

In this Section, we will present the model which will be used in the following to check the applicability of DRs for the TPE contribution to elastic electron-proton scattering. For this purpose, we will evaluate the box graph elastic contribution (corresponding with a nucleon intermediate state) to the structure amplitudes and compare it with the evaluation of the amplitudes using the DR formalism. In our calculation of the box diagram contribution, we will assume an on-shell form of the virtual photon-proton-proton vertex.

\subsection{Loop diagram evaluation}
\label{sec7.1.}

\begin{figure}[htp]
\begin{center}
\includegraphics[width=.65\textwidth]{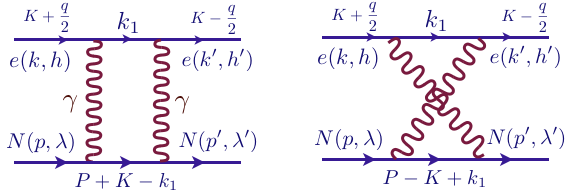}
\end{center}
\caption{Direct and crossed TPE diagrams in $ e^{-} p $ elastic scattering.}
\label{model_graph}
\end{figure}

We will consider the TPE direct and crossed box graph contributions to the structure amplitudes, as shown in Fig.  \ref{model_graph}.
The helicity amplitudes corresponding with the TPE direct and crossed graphs can be expressed  as
\ber
\label{helamp}
 T_{direct}  = & &  - i e^4 \mathop{\mathlarger{\int}}  \frac{  \mathrm{d}^4 k_1}{( 2 \pi )^4} \bar{u}(k',h') \gamma^\mu (\hat{k_1}+m) \gamma^\nu u (k,h) \bar{N}(p',\lambda') \Gamma_\mu (\hat{P} + \hat{K} - \hat{k}_1 + M) \Gamma_\nu N (p,\lambda) \nonumber \\
& & \frac{1}{(k_1 - P - K )^2 - M^2} \frac{1}{k_1^2 - m^2} \frac{1}{(k_1 - K - \frac{q}{2} )^2 - \mu^2 } \frac{1}{(k_1 - K + \frac{q}{2} )^2 - \mu^2 },  \nonumber \\
 T_{crossed}  = & & - i e^4 \mathop{\mathlarger{\int}}  \frac{ \mathrm{d}^4 k_1}{( 2 \pi )^4} \bar{u}(k',h') \gamma^\mu (\hat{k_1}+m) \gamma^\nu u (k,h) \bar{N} (p',\lambda') \Gamma_\nu (\hat{P} - \hat{K} + \hat{k}_1 + M) \Gamma_\mu N (p,\lambda) \nonumber \\
& & \frac{1}{(k_1 + P - K )^2 - M^2} \frac{1}{k_1^2 - m^2} \frac{1}{(k_1 - K - \frac{q}{2} )^2 - \mu^2 } \frac{1}{(k_1 - K + \frac{q}{2} )^2 - \mu^2 },
\eer
where $ \Gamma^\mu $ denotes the virtual photon-proton-proton vertex, $m$ denotes the lepton mass which will be neglected in the following calculations, and where the notation $ \hat{a} \equiv \gamma^\mu a_\mu $ was used. The structure amplitudes entering Eq. (\ref{delta_TPE}) can be expressed as combination of helicity amplitudes with the help of Eq. (\ref{ff_s}).

We perform the box diagram calculation with the assumption of an on-shell form of the virtual photon-proton-proton vertex
\ber
\label{OPE2}
 \Gamma^\mu (Q^2) = \gamma^\mu F_1(Q^2) + \frac{i \sigma^{\mu \nu} q_\nu}{2 M} F_2(Q^2),
\eer
for two models. In the first model the proton is treated as a point particle with charge and anomalous magnetic moment, i.e., the Dirac and Pauli FFs in Eq. (\ref{OPE2}) have the following form
\beq \label{point_model}
 F_1 = 1 , ~~~~ F_2 = \kappa.
 \eeq 
The second model is more realistic and it based on the dipole form for the proton electromagnetic FFs
\ber  \label{dipole_model}
 G_M & = & F_1 + F_2  = \frac{\kappa+1}{( 1 + \frac{Q^2}{\Lambda^2} )^2}, \nonumber \\
 G_E & = &F_1 - \tau F_2  = \frac{1}{( 1 + \frac{Q^2}{\Lambda^2} )^2},
\eer 
with $ \kappa = 1.793 $ and $ \Lambda^2 = 0.71 ~ \mathrm{GeV}^2 $.

\begin{figure}[htp]
\begin{center}
\includegraphics[width=.65\textwidth]{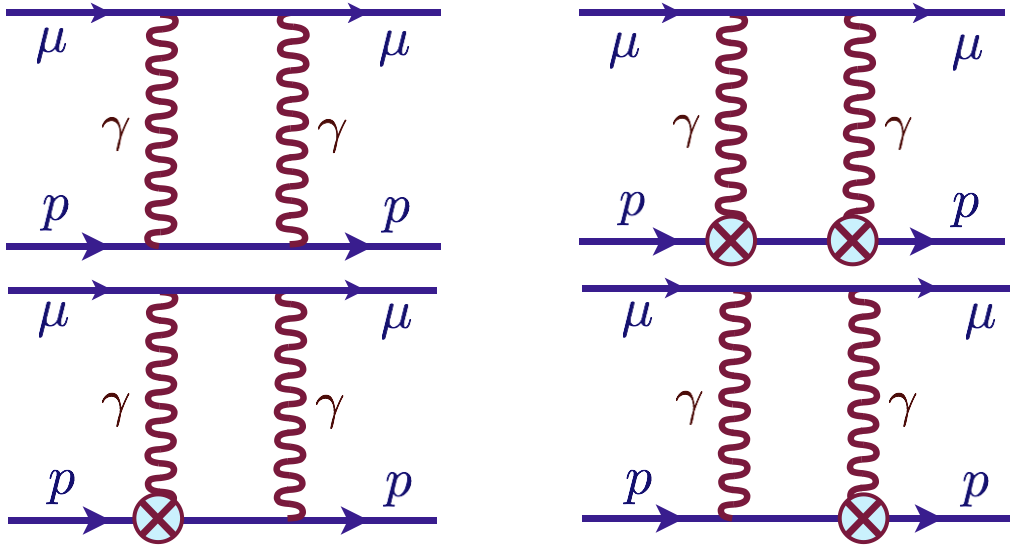}
\end{center}
\caption{The different contributions to the proton box diagram, depending on the different virtual photon-proton-proton vertices. The vertex with (without) the cross denotes the contribution proportional to the $ F_2 $ ($F_1$) FF. The different diagrams show the $ \mathrm{F}_1 \mathrm{F}_1 $ (upper left panel), $ \mathrm{F}_2 \mathrm{F}_2 $ (upper right panel) and $ \mathrm{F}_1 \mathrm{F}_2 $ (lower panels) vertex structures.}
\label{vertices}
\end{figure}

Due to the photon momentum in the numerator of the term proportional to the FF $ F_2 $, the high-energy behavior of the amplitudes will be different depending on whether $ F_1 $ or $ F_2 $ enters the vertex. We denote the contribution with two vector coupling vertices by $ \mathrm{F}_1 \mathrm{F}_1 $, two tensor couplings by $ \mathrm{F}_2 \mathrm{F}_2 $, and the contributions from the mixed case by $ \mathrm{F}_1 \mathrm{F}_2 $, see Fig. \ref{vertices}. We have discussed the HE behavior of the structure amplitudes in case of point-like couplings in Tables I-III. The inclusion of FFs of the dipole form leads to an UV finite results for the structure amplitudes.

We use LOOPTOOLS \cite{Hahn:2000jm,vanOldenborgh:1989wn}  to evaluate the four-point integrals and derivatives of them, as well as to provide a numerical evaluation of the structure amplitudes. The TPE amplitude $ \cG_M $ in the case of scattering of two point charges (i.e., $ \mathrm{F}_1 \mathrm{F}_1 $ contribution with $ F_1 = 1 $) has the IR divergent term 
\beq
 \cG^{IR,~point}_M = \frac{\alpha_{EM}}{\pi} \ln \left(\frac{Q^2}{\mu^2}\right) \left\{ \ln \left(\frac{| u - M^2 |}{s-M^2}\right)+ i \pi \right\},
\eeq
with $ \alpha_{EM} \equiv e^2/4 \pi \simeq 1/137 $. When including FFs, the $ \mathrm{F}_1 \mathrm{F}_1 $ vertex structure gives rise to an IR divergence in the amplitude $ \cG_M $ which is given by  $ \cG_M^{IR, ~F_1F_1} =  F_1(Q^2) \cG^{IR,~point}_M $, whereas the $ \mathrm{F}_1 \mathrm{F}_1 $ contributions to the amplitudes $ \cF_2 $ and $ \cF_3 $ are IR finite. The $ \mathrm{F}_1 \mathrm{F}_2 $ vertex structure gives rise to IR divergences in the amplitude $ \cG_M $ as well as $ \cF_2 $ which are given by $ \cG_M^{IR, ~F_1F_2} = \cF_2^{IR, ~F_1F_2} = F_2(Q^2) \cG^{IR,~point}_M $,
whereas the $ \mathrm{F}_1 \mathrm{F}_2 $ contribution to the amplitude $ \cF_3 $ is IR finite. Finally, the $ \mathrm{F}_2 \mathrm{F}_2 $ vertex structure contribution to all these amplitudes is IR finite. When combining all IR divergent pieces, Eq. (\ref{delta_TPE}) yields the IR divergent TPE correction
\ber \label{IR_delta_TPE}
\delta^{IR}_{2 \gamma} = \frac{2 \alpha_{EM}}{\pi} \ln \left(\frac{Q^2}{\mu^2}\right) \ln \left(\frac{| u - M^2 |}{s-M^2}\right).
\eer
When comparing with data, which are radiatively corrected, we subtract Eq.~(\ref{IR_delta_TPE}) in the cross section formula of Eq.~(\ref{delta_TPE}). This is in agreement with the Maximon and Tjon (MaTj) prescription for the soft photon TPE contribution, i.e., $ \delta^{\mathrm{MaTj}}_{2 \gamma, ~\mathrm{soft}} = \delta^{IR}_{2 \gamma} $, see Eq.~(3.39) of Ref.~\cite{Maximon:2000hm}. Note that the $ P_t $ and $ P_l $ observables, Eqs.~(\ref{polarization_observables1},\ref{polarization_observables2}), are free of IR divergencies.

\subsection{Dispersive evaluation}
\label{sec7.2.}

We next discuss the evaluation of the box diagram contributions with nucleon intermediate states using DRs. We perform DR calculations separately for $ \mathrm{F}_1 \mathrm{F}_1 $, $ \mathrm{F}_1 \mathrm{F}_2 $ and $ \mathrm{F}_2 \mathrm{F}_2 $ vertex structures (Fig. \ref{vertices}) for both FF models described above in Eqs. (\ref{point_model}, \ref{dipole_model}).

For the point-like model, we obtain analytical expressions for the imaginary part of the structure amplitudes. The imaginary parts of the structure amplitudes due to the $ \mathrm{F}_1 \mathrm{F}_1 $ vertex structure are given by
\ber
\label{imF1F1point}
 \Im \cG_M & = & \alpha_{EM} \left\{ \ln\left( \frac{Q^2}{\mu^2}\right) - \frac{s+M^2}{2s} - \frac{2\left(s-M^2\right)^2- s Q^2}{2\left(\left(s-M^2\right)^2 - s Q^2\right)}\ln \left(\frac{ s Q^2}{\left(s-M^2\right)^2}\right) \right\},  \nonumber \\
 \Im \cF_2 & = & \frac{\alpha_{EM} M^2 Q^2}{\left(s-M^2\right)^2 - s Q^2}  \left\{ 1 + \frac{\left(s-M^2\right)^2}{\left(s-M^2\right)^2 - s Q^2}\ln \left(\frac{ s Q^2}{\left(s-M^2\right)^2}\right)\right\}, \nonumber  \\
 \Im F_3 & = & \frac{\alpha_{EM} M^2 \left( s - M^2 \right)}{\left(s-M^2\right)^2  - s Q^2} \left\{   \frac{s+M^2}{s} + \frac{\left(s-M^2\right)\left(2\left(s-M^2\right) - Q^2\right)}{\left(s-M^2\right)^2 - s Q^2}\ln (\frac{ s Q^2}{\left(s-M^2\right)^2})\right\}.
\eer
The imaginary parts of the structure amplitudes due to the mixed $ \mathrm{F}_1 \mathrm{F}_2 $ vertex structure are given by
\ber
\label{imF1F2point}
\Im \cG_M & = & \alpha_{EM} \kappa \left\{  \ln\left( \frac{Q^2}{\mu^2}\right)  - \frac{M^2}{s} + \frac{2\left(s-M^2\right)^2 - s Q^2}{2\left(\left(s-M^2\right)^2 - s Q^2\right)}\ln \left(\frac{ s Q^2}{\left(s-M^2\right)^2}\right) \right\},   \nonumber \\
\Im \cF_2 & = & \alpha_{EM} \kappa \frac{M^2 Q^2}{ \left(s-M^2\right)^2 - s Q^2} \left\{ 1 + \frac{\left(s-M^2\right)^2\left(s+2M^2\right) - s^2 Q^2}{2 M^2 \left(\left(s-M^2\right)^2 - s Q^2\right)} \ln \left(\frac{ s Q^2}{\left(s-M^2\right)^2}\right) \right\}   \nonumber \\ 
& &  + \alpha_{EM} \kappa \ln\left( \frac{Q^2}{\mu^2} \right) ,  \nonumber \\
\Im \cF_3 & = & \alpha_{EM} \kappa \frac{ M^2}{s \left( \left(s-M^2\right)^2 - s Q^2 \right)}  \Bigg\{ 2M^2\left(s-M^2\right) + s Q^2  \nonumber \\
 & & + \frac{ s \left(s-M^2\right)^2\left(2\left(s-M^2\right) - Q^2\right)}{\left(s-M^2\right)^2 - s Q^2 }  \ln \left(\frac{ s Q^2}{\left(s-M^2\right)^2}\right) \Bigg\}.     
\eer
The imaginary parts of the structure amplitudes due to the $ \mathrm{F}_2 \mathrm{F}_2 $ vertex structure are given by 
\ber
\label{imF2F2point}
 \Im \cG_M & = & \alpha_{EM} \kappa^2 \frac{s-M^2}{2s} \left\{ 1 +  \frac{s^2 Q^2}{4M^2 \left(\left(s-M^2\right)^2 - s Q^2\right)}\ln \left(\frac{ s Q^2}{\left(s-M^2 \right)^2}\right) \right\},  \nonumber \\
 \Im \cF_2 & = & - \frac{\alpha_{EM} \kappa^2}{4} \frac{\left(s-M^2\right) Q^2}{\left(s-M^2\right)^2 - s Q^2} \left\{ 1+ \frac{\left(s-M^2\right)^2}{ \left(s-M^2\right)^2 - s Q^2}\ln \left(\frac{ s Q^2}{\left(s-M^2\right)^2}\right)\right\},  \nonumber \\
 \Im \cF_3 & = & - \frac{\alpha_{EM} \kappa^2}{4 s \left(  \left(s-M^2\right)^2 - s Q^2 \right)} \Bigg\{ 4 M^2 \left(s-M^2\right)^2 + s Q^2 \left(s - 3 M^2\right) \nonumber \\
 & &  +  \frac{\left(M^6 - 3 M^2 s^2 + 2 s^3 - s^2 Q^2 \right) s Q^2}{\left(s-M^2\right)^2 - s Q^2}\ln \left(\frac{ s Q^2}{\left(s-M^2\right)^2}\right) \Bigg\}.
\eer

We checked that the numerical calculations of the imaginary part of the structure amplitudes are in agreement with predictions for the target normal spin asymmetry $ A_n $ \cite{Pasquini:2004pv} for the model with dipole form of electromagnetic FFs \cite{DeRujula:1972te}.

In Section~\ref{sec8}, we will compare the DR and the direct loop evaluations for the TPE contribution which results from the nucleon intermediate state. 

\subsection{Forward limit}
\label{sec7.3.}

Before presenting the numerical results for the TPE corrections at finite $Q^2$, 
we first discuss the forward limit. This limit is relevant to extract the proton charge radius 
from elastic scattering data. 
In the forward limit, corresponding with $ Q^2 \to 0 $ and $ \varepsilon \to 1 $, the TPE correction to the cross section is given by Coulomb photons from the $ \mathrm{F}_1 \mathrm{F}_1 $ structure of virtual photon-proton-proton vertices. This result was first obtained for the electron-proton scattering in Dirac theory as the first order cross section correction by McKinley and Feshbach \cite{McKinley:1948zz}. The so-called Feshbach correction to the cross section can be expressed analytically in terms of the laboratory scattering angle $ \theta $ or the photon polarization parameter $ \varepsilon $ as
\ber \label{Feshbach_term}
\delta_{F} = \pi \alpha_{EM} \frac{\sin \frac{\theta}{2} - \sin^2 \frac{\theta}{2}}{\cos^2 \frac{\theta}{2}} \approx  \pi \alpha_{EM} \frac{\sqrt{1-\varepsilon}}{\sqrt{1-\varepsilon}+\sqrt{1+\varepsilon}}.
\eer

It is instructive to provide some analytical expressions for $ \delta_{2 \gamma} $ in the forward limit resulting from the $ \mathrm{F}_1 \mathrm{F}_1 $ vertex contribution to the full box diagram calculation. 

For the case of electron scattering off massless quarks (taken with unit charge) the TPE correction is given by \cite{Afanasev:2005mp}
\ber
\delta_{2 \gamma}  =  \frac{\alpha_{EM}}{\pi} \Bigg\{ 2 \ln \left( \frac{Q^2}{\mu^2} \right) \ln \left( \frac{1-x}{1+x}\right)  &+ & \frac{x}{1+x^2} \left[ \ln^2 \left( \frac{1+x}{2x} \right) + \ln^2 \left( \frac{1 - x}{2x} \right) + \pi^2  \right] \nonumber \\
& - &  \frac{x}{1+x^2} \left[ \ln \left( \frac{1-x^2}{4x^2} \right) - x \ln \left( \frac{1 + x}{ 1 - x } \right) \right] \Bigg\},
\eer
with $ x = \sqrt{1 - \varepsilon} / \sqrt{1 + \varepsilon} $ and $ Q^2 = 4 x \nu $. In the forward limit ($Q^2 \to 0$ and $ \varepsilon \to 1 $, at finite $ \nu $) we recover the Feshbach term and find large logarithmic correction terms in $(1-\varepsilon) $~
\ber \label{low_quark_mass}
\delta_{2 \gamma} -  \delta^{IR}_{2 \gamma} & \longrightarrow & \delta_F + \frac{\alpha_{EM}}{\pi} \sqrt{\frac{1-\varepsilon}{2}} \ln\left( 2 \left(1-\varepsilon \right) \right) \left[ \frac{1}{2} \ln\left(2 \left(1-\varepsilon \right)\right) + 1 \right],
\eer
where the IR divergent TPE is given by the massless limit of Eq.~(\ref{IR_delta_TPE}). 

For the case of forward scattering off a massive point particle we also give the analytical 
form of the momentum transfer expansion of the $ \mathrm{F}_1 \mathrm{F}_1 $ vertex contribution for the model with point particles. In the forward direction, only the amplitude $\cG_2$ defined in Eq.~(\ref{new_amplitude2}) survives since $\delta_{2 \gamma} \to 2 \Re \cG_2$ in the forward limit. The $ \mathrm{F}_1 \mathrm{F}_1 $ point vertex contribution to the imaginary part of $\cG_2$ is obtained from Eqs.~(\ref{imF1F1point}) as~
\ber
\label{imG2}
 \Im \cG_2^{F_1 F_1}  & = & \alpha_{EM} \left\{ \ln\left( \frac{Q^2}{\mu^2}\right) + \frac{Q^2}{4s} + \frac{Q^2}{8} \frac{s}{\nu^2 - \nu_{ph}^2} \ln \left(\frac{ s Q^2}{\left(s-M^2\right)^2} \right) \right\},  
 \eer
with $\nu_{ph}$ as defined in Eq.~(\ref{nuphys}). 
Using the dispersion relation of Eq.~(\ref{oddDR}), we can express the real part of $\cG_2$  in the forward limit (for $Q^2 \ll M^2$) in terms of $Q^2$ and the electron beam energy $ E_e^{lab} $ in the laboratory frame as~
\ber
\label{reG2}
&& \Re \cG_2^{F_1 F_1}   \longrightarrow \frac{\alpha_{EM}}{\pi} \left\{ -  \frac{Q^2}{2 M E_e^{lab}} \ln \left( \frac{Q^2}{\mu^2} \right) 
+  \pi^2 \frac{Q}{4 E_e^{lab}} \right. \nonumber \\
&&\hspace{3cm} + \left. 
 \frac{Q^2}{2 M E_e^{lab}} 
\ln \left( \frac{Q}{2 E_e^{lab}} \right) \left[ \ln \left( \frac{Q}{2 E_e^{lab}} \right) + 1 \right] + \mathrm{O} \left(\frac{Q^2}{M^2} \right) \right\} , 
\eer 
where we have dropped terms of order $Q^2 / M^2$ which do not lead to any logarithmic enhancements in the near forward direction. 
Note that we can equivalently express Eq.~(\ref{reG2}) through the variable $\varepsilon$ using the kinematical relation $Q / E_e^{lab} \simeq \sqrt{2 (1 - \varepsilon)}$, which holds in the 
forward direction. 
Eq.~(\ref{reG2}) then allows to directly express $\delta_{2 \gamma}$ in the forward direction as~
\ber
\delta_{2 \gamma} -  \delta^{IR}_{2 \gamma} & \longrightarrow & \delta_F  +  \frac{\alpha_{EM}}{\pi} 
 \frac{Q^2}{M E_e^{lab}} 
\ln \left( \frac{Q}{2 E_e^{lab}} \right) \left[ \ln \left( \frac{Q}{2 E_e^{lab}} \right) + 1 \right] + \mathrm{O} \left(\frac{Q^2}{M^2} \right), 
\label{deltalogcorr}
\eer
where the leading finite term (proportional to $Q / E_e^{lab}$) is obtained as the Feshbach correction term $\delta_F$, and where subleading logarithmic correction terms are also shown.  We found that our forward limit result of Eq.~(\ref{deltalogcorr}) agrees with an expression obtained some time ago~\cite{Brown:1970te}~\footnote{Note that in Ref.\cite{Gorchtein:2014hla}, the finite logarithmic terms multiplying $Q^2$ were missed. 
Eq.~(\ref{deltalogcorr}) shows that the elastic box contains terms proportional to $Q^2 \ln(Q^2)$ and $Q^2 \ln^2(Q^2)$, which lead to corrections in the near forward direction.}. 

We can similarly study the contributions of the $ \mathrm{F}_1 \mathrm{F}_2 $ and $ \mathrm{F}_2 \mathrm{F}_2 $ vertex structures to the amplitude $ \cG_2 $ in the case of a point-like proton comparing their imaginary parts $ \Im \cG_2^{F_1 F_2}$, and $\Im \cG_2^{F_2 F_2} $ with the expression for the $ \mathrm{F}_1 \mathrm{F}_1 $ vertex structure $ \Im \cG_2^{F_1 F_1} $ of Eq. (\ref{imG2}). The corresponding imaginary parts can be obtained from Eqs.~(\ref{imF1F2point}, \ref{imF2F2point}) as
\ber
\label{imG2F1F2}
 \Im \cG_2^{F_1 F_2} & = & \alpha_{EM} \frac{Q^2}{4 M^2}  \kappa \left\{ - \ln\left( \frac{Q^2}{\mu^2}\right) + \frac{2 M^2}{s} - \frac{Q^2}{8} \frac{s}{\nu^2 - \nu_{ph}^2} \ln \left(\frac{ s Q^2}{\left(s-M^2\right)^2} \right) \right\},   \\
\Im \cG_2^{F_2 F_2}  & = & \alpha_{EM} \frac{Q^2}{4M^2} \kappa^2  \left\{  - \frac{ s - 2 M^2}{2 s}  -\frac{Q^2}{16} \frac{s}{\nu^2 - \nu_{ph}^2} \ln \left(\frac{ s Q^2}{\left(s-M^2\right)^2} \right) - \frac{1}{2} \ln \left(\frac{ s Q^2}{\left(s-M^2\right)^2} \right) \right\}.  \nonumber \\
\label{imG2F2F2}
\eer
The Feshbach correction and the subleading logarithmic terms in the real part of the amplitude $ \cG_2^{F_1 F_1}  $, Eq.~(\ref{reG2}), arise from the logarithmic term in Eq.~(\ref{imG2}). 
Analogous terms are suppressed by the pre-factor $ Q^2 / M^2 $ in the imaginary parts for the $ \mathrm{F}_1 \mathrm{F}_2 $ and $ \mathrm{F}_2 \mathrm{F}_2 $ vertex structures in comparison with the $ \mathrm{F}_1 \mathrm{F}_1 $ vertex structure. The additional logarithmic term in Eq.~(\ref{imG2F2F2}) also leads to corrections of higher order in $ Q / M $ in comparison with 
Eq.~(\ref{reG2}).
Besides the elastic contribution discussed here, 
Ref.~\cite{Brown:1970te} also derived that in the forward limit, the 
$Q^2 \ln(Q/2 E_e^{lab})$ term in Eq.~(\ref{deltalogcorr}) obtains an additonal contribution 
due to inelastic states, which can be expressed through the total photo-production cross section on a nucleon.

\begin{figure}[htp]
\begin{center}
\includegraphics[width=1.\textwidth]{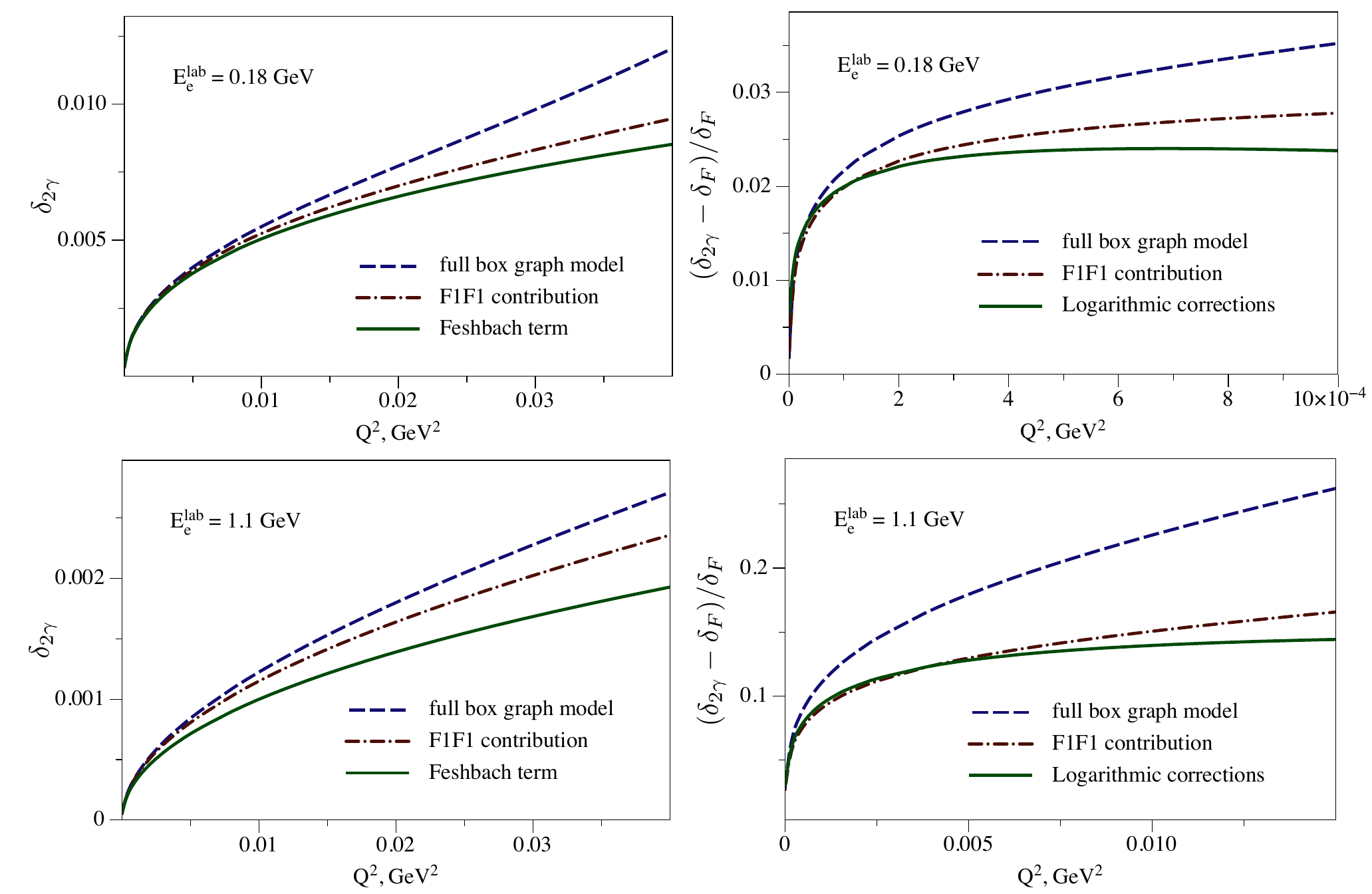}
\end{center}
\caption{The small $ Q^2 $-limit of the TPE correction in the model with a point-like proton for $ E^{lab}_e = 0.18 ~\mathrm{GeV} $ (upper panels) and $ E^{lab}_e = 1.1 ~\mathrm{GeV} $ (lower panels). For clarity, the contribution relative to the Feshbach term is shown on the right panels  
for the logarithmic correction term of Eq.~(\ref{deltalogcorr}), for the F$_1$F$_1$ vertex contribution to the box diagram, and for the full box diagram calculation, also including the F$_1$F$_2$ and F$_2$F$_2$ contributions.}
\label{delta_vs_Q2_LE1} 
\end{figure}

In Fig.~\ref{delta_vs_Q2_LE1}, we compare the Feshbach correction with the full box diagram calculation of $ \delta_{2 \gamma}$ for point protons at low momentum transfers and at beam energies corresponding with experiments at MAMI and JLab. One sees that at small $Q^2$, the leading TPE contribution is given by the $ \mathrm{F}_1 \mathrm{F}_1 $ vertex structure, and approaches the Feshbach term in the forward direction. We furthermore see that at small $Q^2$, the leading corrections to the Feshbach result are given by the 
logarithmic terms given in Eq.~(\ref{deltalogcorr}). 

\begin{figure}[htp]
\begin{center}
\includegraphics[width=1.0\textwidth]{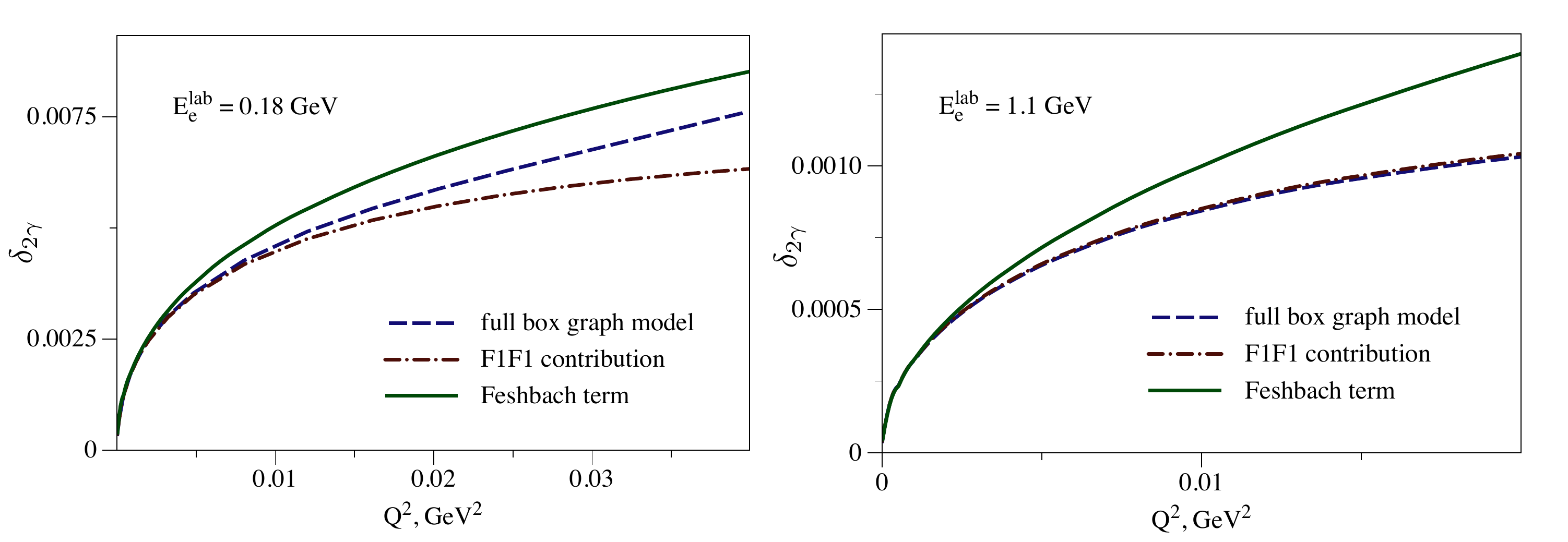}
\end{center}
\caption{The small $ Q^2 $-limit of the TPE correction in the model with dipole proton FFs for $ E^{lab}_e = 0.18 ~\mathrm{GeV} $ (left panel) and $ E^{lab}_e = 1.1 ~\mathrm{GeV} $ (right panel). }
\label{delta_vs_Q2_LE_dipole}
\end{figure}

In Fig.~\ref{delta_vs_Q2_LE_dipole}, we compare the analogous results using the dipole model for the proton FFs. 

\section{Results and discussion}
\label{sec8}

In this Section, we firstly compare the model calculation of the elastic contribution to TPE amplitudes with the evaluation within the DR formalism. 
Subsequently, we discuss predictions for unpolarized and polarization transfer observables 
of elastic electron-proton scattering and compare with existing data. 

\begin{figure}[htp]
\begin{center}
\includegraphics[width=1.\textwidth]{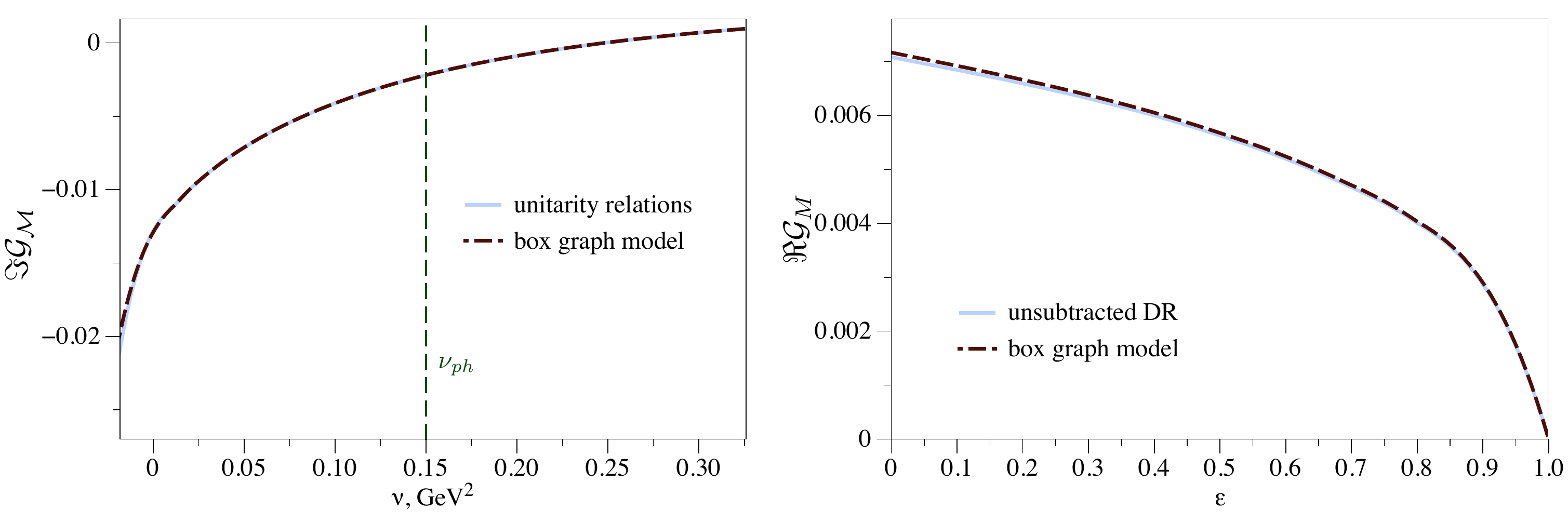}
\end{center}
\caption{ Imaginary part (left panel) and real part (right panel) of the structure amplitude $ \cG_M $ for the $ \mathrm{F}_1 \mathrm{F}_1 $ vertex structure with dipole FFs for $ Q^2 = 0.1 ~\mathrm{GeV}^2 $. The vertical line in the left panel corresponds with the boundary between physical and unphysical regions, i.e., $ \nu_{ph} = 0.15 ~\mathrm{GeV}^2 $.}
\label{amplitude_gm_f1f1}
\end{figure}

\begin{figure}[htp]
\begin{center}
\includegraphics[width=1.\textwidth]{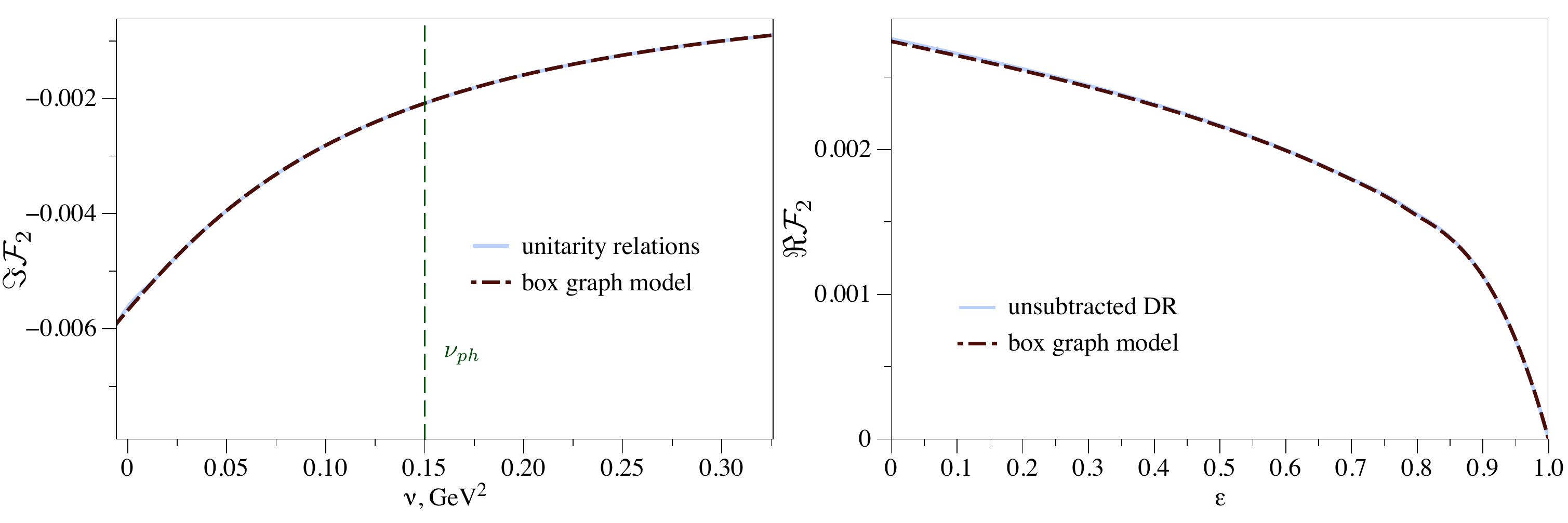}
\end{center}
\caption{ Same as Fig. \ref{amplitude_gm_f1f1}, but for the structure amplitude $ \cF_2 $.}
\label{amplitude_f2_f1f1}
\end{figure}

\begin{figure}[htp]
\begin{center}
\includegraphics[width=1.\textwidth]{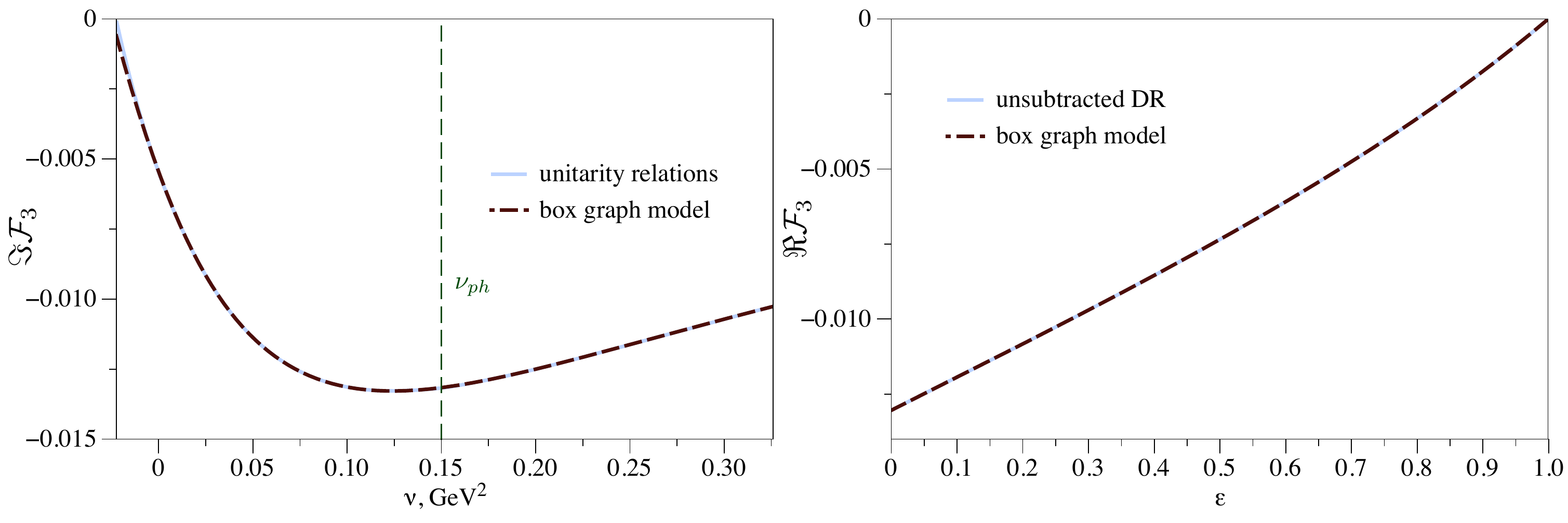}
\end{center}
\caption{ Same as Fig. \ref{amplitude_gm_f1f1}, but for the structure amplitude $ \cF_3 $.}
\label{amplitude_f3_f1f1}
\end{figure}

The results for the real and imaginary parts of the amplitudes for the case of the $ \mathrm{F}_1 \mathrm{F}_1 $ vertex structure in the model with dipole FFs are shown in Figs. \ref{amplitude_gm_f1f1} -  \ref{amplitude_f3_f1f1}. We show the unitarity relations calculation of the imaginary parts of the structure amplitudes both in physical and unphysical regions. For the latter, we use the analytical continuation as outlined in Section \ref{sec3}. For the imaginary parts, we see a perfect agreement between the unitarity relations calculations and the box graph evaluation both in physical and unphysical regions. We also checked that for the  $ \mathrm{F}_1 \mathrm{F}_2$  and $ \mathrm{F}_2 \mathrm{F}_2$ vertex structures the imaginary parts of the structure amplitudes are in perfect agreement between the two approaches for all amplitudes and for both FF models. This is to be expected as the imaginary parts of the structure amplitudes correspond with an intermediate state in the box diagram which is on its mass shell. Therefore only on-shell information enters the imaginary parts.

\begin{figure}[htp]
\begin{center}
\includegraphics[width=1.\textwidth]{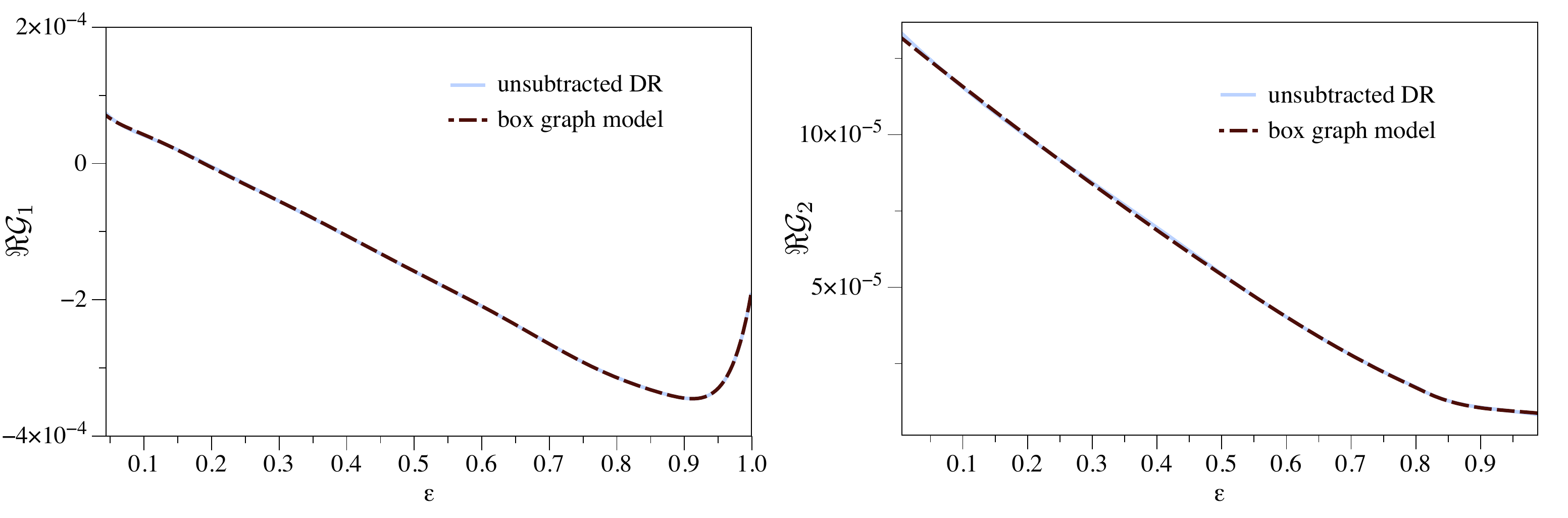}
\end{center}
\caption{$ \varepsilon $-dependence of the real part of the structure amplitudes $ \cG_1, \cG_2 $  in case of the $ \mathrm{F}_2 \mathrm{F}_2 $ vertex structure with dipole FFs for $ Q^2 = 0.1 ~\mathrm{GeV}^2 $.}
\label{g1g2_unsubtracted}
\end{figure}

\begin{figure}[htp]
\begin{center}
\includegraphics[width=1.\textwidth]{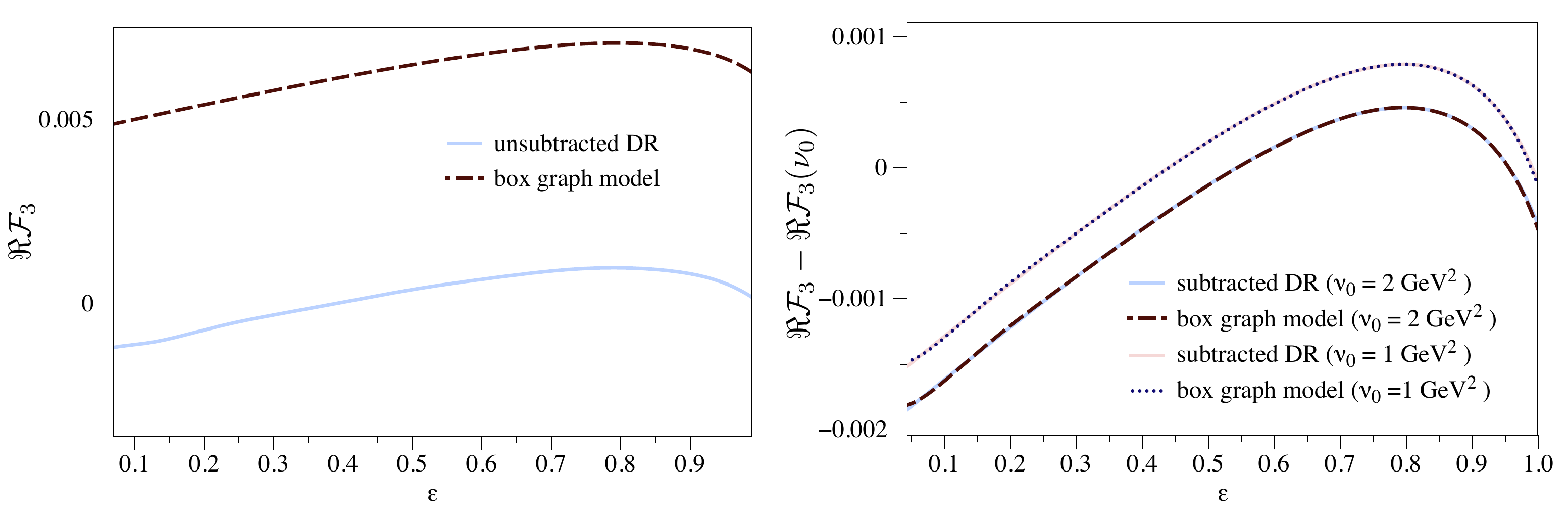}
\end{center}
\caption{$ \varepsilon $-dependence of the real part of the structure amplitude $ \cF_3 $ in case of the $ \mathrm{F}_2 \mathrm{F}_2 $ vertex structure with dipole FFs for $ Q^2 = 0.1 ~\mathrm{GeV}^2 $. Left panel: comparison of the box diagram evaluation with unsubtracted DR. Right panel: comparison between the box diagram and DR evaluations when performing one subtraction. The calculations are shown for two different subtraction points: 
$ \nu_0 = 1 ~\mathrm{GeV}^2$, and $\nu_0 = 2 ~\mathrm{GeV}^2 $.}
\label{f3_subtracted}
\end{figure}

For the real parts, we use the unsubtracted DRs at fixed $ Q^2 $. By comparing the DR results with the loop diagram evaluation for $ \mathrm{F}_1 \mathrm{F}_1 $ vertex structure of the real parts (for the sum of direct and crossed box diagrams), we see from Figs.~\ref{amplitude_gm_f1f1} - \ref{amplitude_f3_f1f1} that they nicely agree over the whole physical region of the parameter $ \varepsilon $, which is related to $ \nu $ as 
\ber
\nu = \sqrt{\frac{1+\varepsilon}{1-\varepsilon}} \nu_{ph},
\eer
with $ \nu_{ph}$ defined in Eq. (\ref{nuphys}). We checked that in case of the $ \mathrm{F}_1 \mathrm{F}_2 $ vertex structure, the real parts as obtained from the box diagram model calculation also agree with the unsubtracted DR results. In case of the $ \mathrm{F}_2 \mathrm{F}_2 $ vertex structure, the unsubtracted DRs reproduce the box diagram model results for the amplitudes $ \cF_2, \cG_1, \cG_2 $  for both FF models. As an example, we show the results for $\cG_1$ and $\cG_2$ in Fig. \ref{g1g2_unsubtracted}. These amplitudes are UV finite in case of the point-like model calculation. The real part of the $ \cF_3$ amplitude requires an UV regularization for the point-like box graph model. Consequently the DR for the amplitude $ \cF_3 $ requires one subtraction. The resulting subtraction term cannot be reconstructed from the imaginary part of the amplitude $ \cF_3 $. 
 This term describes the contribution of physics at high energies to low-energy processes. 
When using dipole FFs for the $ \mathrm{F}_2 \mathrm{F}_2 $ vertex structure, 
one finds that the unsubtracted DR for the elastic contributions also converges for $ \cF_3 $. 
The results for the real part of the structure amplitude $ \cF_3 $ for the case of the $ \mathrm{F}_2 \mathrm{F}_2 $ vertex structure are shown in Fig.~\ref{f3_subtracted}. 
One firstly noticed from Fig.~\ref{f3_subtracted} (left panel) that the calculated real part 
of $ \cF_3 $ in the box graph model does not agree with the 
amplitude reconstructed using unsubtracted DRs. 
Although the box diagram calculation for $ \cF_3 $ is convergent for the $ \mathrm{F}_2 \mathrm{F}_2 $ vertex structure when using on-shell vertices with dipole FFs, we like to stress that this result is model dependent. We notice however that after performing one subtraction, we find an agreement between the DR calculation and the box diagram model evaluation, see right panel of Fig.~\ref{f3_subtracted}.
Even though numerically the Feynman diagram calculation may yield satisfactory results 
over some kinematic range, as will be shown in the following, fixing the subtraction function to reproduce the Feynman diagram calculation with effective vertices would be a model dependent assumption, and is not a consequence of quantum field theory. As a first step, we will fix the subtraction function in the following to empirical TPE data, with the assumption of  only the nucleon intermediate state contribution. A  fully consistent application of the DR formalism will require also to add the inelastic term, and then fit the subtraction term to the data. Such inclusion of inelastic states is beyond the scope of the present work.

\begin{figure}[htp]
\begin{center}
\includegraphics[width=1.\textwidth]{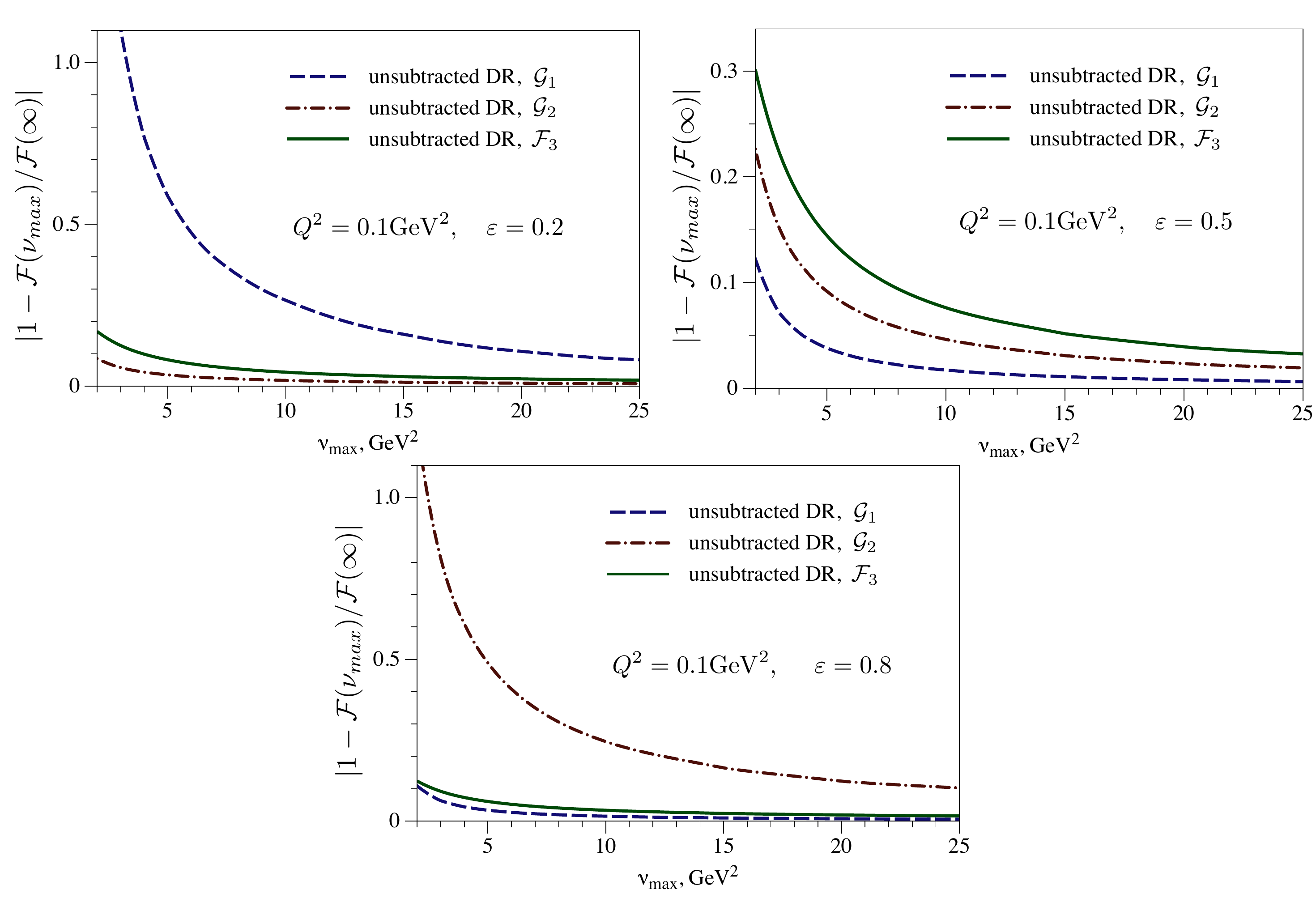}
\end{center}
\caption{Real parts of $\cG_1 $, $\cG_2 $, and $\cF_3 $ evaluated through unsubtracted DRs, as function of the upper integration limit $\nu_{max}$. The plot shows the relative deviation for each amplitude from its value for $\nu_{max} = \infty$, denoted by $\cF(\infty)$, where $\cF$ stands for $\cG_1, \cG_2, \cF_3$.   
All results are for the $ \mathrm{F}_2 \mathrm{F}_2 $ vertex structure with dipole FFs.}
\label{conv_test}
\end{figure}

To test the numerical convergence for different kinematical situations, we show in Fig.~\ref{conv_test} the contributions to the real parts of $\cG_1 $, $\cG_2 $, and $\cF_3 $ evaluated through unsubtracted DRs, as function of the upper integration limit in the DRs. We see from Fig.~\ref{conv_test} that for the case of the $ \mathrm{F}_2 \mathrm{F}_2 $ vertex structure with dipole FFs , the convergence of unsubtracted DRs is slowest at large (small) values of $ \varepsilon $ for $ \cG_2 ~(\cG_1) $ respectively, while at intermediate values of $ \varepsilon $ the slowest convergence occurs for $\cF_3 $. For a phenomenological evaluation of the TPE contribution to elastic electron-nucleon scattering, we like to minimize any model dependence due to higher 
energy contributions. In a full calculation, such contributions arise from inelastic states which always will require some approximate treatment. To minimize any such uncertainties and to provide a more flexible formalism when applied to data, we propose to consider a DR formalism with one subtraction for the amplitude $\cF_3 $. The subtraction constant will be obtained by a fit to elastic electron-nucleon scattering observables, in the region where precise data are available. 

We next discuss the implementation of such a subtracted DR formalism for the TPE contribution and provide a detailed comparison to different observables. The TPE correction to the unpolarized elastic electron-proton scattering cross section in Eq. (\ref{unpolarized_cross_section}) can be expressed as the sum of a term evaluated using an unsubtracted DR and a term arising from the $ \mathrm{F}_2 \mathrm{F}_2 $ contribution of $ {\cal{F}}_3 $, which we will evaluate by performing a subtraction~:
\ber
 \delta_{2 \gamma} & = & \delta^{0}_{2 \gamma} + f\left(\nu,Q^2 \right) \Re {\cal{F}}^{F_2F_2}_3,
\eer
with
\ber
 \delta^0_{2 \gamma} & = & \frac{1}{G^2_M  + \frac{\varepsilon}{\tau} G^2_E } \left( 2 G_M  \Re  {\cal{G}}_1 +2 \frac{\varepsilon}{\tau} G_E  \Re  {\cal{G}}_2 + 2 G_M \left(\varepsilon -1 \right) \frac{\nu}{M^2} \Re {\cal{F}}^{F_1F_1+F_1F_2}_3 \right),
\eer
and
\ber \label{coefficients_subtracted}
f(\nu,Q^2) & = & \frac{2 {G_M} (\varepsilon -1 ) }{G^2_M  + \frac{\varepsilon}{\tau} G^2_E }  \frac{\nu}{M^2}.
\eer

The polarization transfer observables of Eqs. (\ref{polarization_observables1}, \ref{polarization_observables2}) can also be expressed as model independent terms $ (\frac{P_t}{P_l})^0  $, $ (\frac{P_l}{P^{Born}_l})^0 $ and the contribution due to $ {\cal{F}}^{F_2F_2}_3 $  as
\ber 
 \frac{P_t}{P_l} & = & \left(\frac{P_t}{P_l} \right)^0 + g(\nu,Q^2) \Re \cF^{F_2F_2}_3,  \label{ptplsub} \\ 
 \frac{P_l}{P^{Born}_l} & = & \left(\frac{P_l}{P^{Born}_l} \right)^0 + h(\nu,Q^2) \Re \cF^{F_2F_2}_3, \label{plplsub}
\eer
with
\ber \label{coefficients_subtracted2}
g(\nu,Q^2) & = & - \sqrt{\frac{2 \varepsilon}{\tau ( 1 + \varepsilon )}} \frac{ 1 - \varepsilon }{ 1 + \varepsilon } \frac{G_E}{G_M^2} \frac{\nu}{M^2},  \\
 h(\nu,Q^2) & = & - \frac{2 \varepsilon}{\tau G^2_M + \varepsilon G^2_E} \frac{1}{G_M} \frac{ \varepsilon \tau G^2_M + G^2_E }{ 1 + \varepsilon } \frac{\nu}{M^2} . 
\eer

\begin{figure}[htp]
\begin{center}
\includegraphics[width=.65\textwidth]{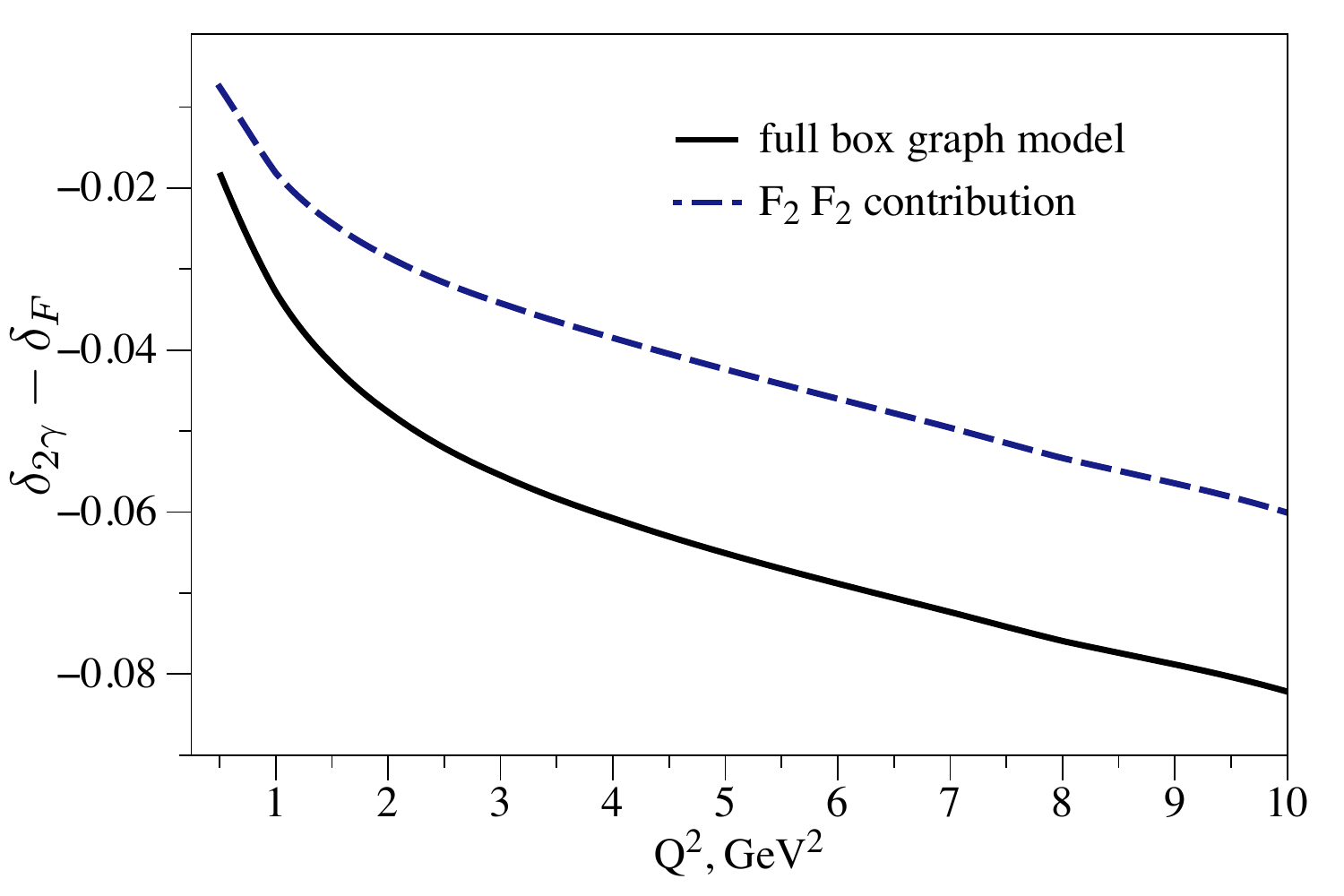}
\end{center}
\caption{ Model prediction for the TPE correction  $ \delta_{2 \gamma} -\delta_F $, with $ \delta_F $ is the Feshbach term of Eq. (\ref{Feshbach_term}), for $ \varepsilon = 0.01 $.}
\label{delta_vs_Q2}
\end{figure}

The predictions for the elastic electron-proton scattering observables can be made with one subtraction point at $ \nu = \nu_0 $, which we express as
\ber
 \delta_{2 \gamma}(\nu,Q^2) & = & \delta^{0}_{2 \gamma}(\nu,Q^2) + f(\nu,Q^2) \left[ \Re {\cal{F}}^{F_2F_2}_3 (\nu,Q^2) - \Re {\cal{F}}^{F_2F_2}_3 (\nu_0,Q^2) \right] \nonumber \\
 & + & f(\nu,Q^2) \, \Re {\cal{F}}^{F_2F_2}_3 (\nu_0,Q^2) 
 \label{delta_prediction}, 
 \eer
 where we can express the subtraction function $ \Re {\cal{F}}^{F_2F_2}_3 (\nu_0,Q^2) $ through $  \delta_{2 \gamma}(\nu_0,Q^2) $, which has to be obtained from experiment, as
 \ber \label{subtraction_term}
 \Re {\cal{F}}^{F_2F_2}_3 (\nu_0,Q^2) = \frac{\delta_{2 \gamma}(\nu_0,Q^2) - \delta^{0}_{2 \gamma}(\nu_0,Q^2) }{f(\nu_0,Q^2)}. 
 \eer
 We can then insert this subtraction term (for every fixed value of $ Q^2 $) into Eqs. (\ref{ptplsub}, \ref{plplsub}) and make predictions for the $ \nu $ or $ \epsilon $ dependence of these observables as
  \ber
 \left(\frac{P_t}{P_l} \right)(\nu,Q^2) & = & \left(\frac{P_t}{P_l} \right)^{0}(\nu,Q^2) + g(\nu,Q^2)\left[ \Re {\cal{F}}^{F_2F_2}_3 (\nu,Q^2) - \Re {\cal{F}}^{F_2F_2}_3 (\nu_0,Q^2) \right] \nonumber \\
 & + & g(\nu,Q^2) \frac{\delta_{2 \gamma}(\nu_0,Q^2) - \delta^{0}_{2 \gamma}(\nu_0,Q^2) }{f(\nu_0,Q^2)} ,   \label{ptpl_prediction} \\
 \left(\frac{P_l}{P^{Born}_l}\right)(\nu,Q^2) & = & \left(\frac{P_l}{P^{Born}_l} \right)^{0}(\nu,Q^2) + h(\nu,Q^2)\left[ \Re {\cal{F}}^{F_2F_2}_3 (\nu,Q^2) - \Re {\cal{F}}^{F_2F_2}_3 (\nu_0,Q^2) \right]\nonumber \\
 & + & h(\nu,Q^2) \frac{\delta_{2 \gamma}(\nu_0,Q^2) - \delta^{0}_{2 \gamma}(\nu_0,Q^2) }{f(\nu_0,Q^2)}. \label{plpl_prediction}
\eer
In Eqs. (\ref{delta_prediction}, \ref{ptpl_prediction}, \ref{plpl_prediction}) the difference $ \Re {\cal{F}}^{F_2F_2}_3 (\nu,Q^2) - \Re {\cal{F}}^{F_2F_2}_3 (\nu_0,Q^2) $ is calculated from a subtracted DR. In the following, we determine the subtraction term from the unpolarized cross section measurements \cite{Bernauer:2013tpr}, and show our predictions for the different observables. The TPE correction to the unpolarized elastic electron-proton scattering evaluated in the model calculation of Section \ref{sec7}, with the Feshbach term subtracted, is shown in Fig. \ref{delta_vs_Q2} for a small value of $ \varepsilon $. 
It is seen from Fig. \ref{delta_vs_Q2} that the departure of the TPE correction from the Feshbach term strongly increases with increasing $Q^2$. One also sees that at larger $Q^2$, this is mainly due to the contribution from the $ \mathrm{F}_2 \mathrm{F}_2$ vertex structure. 

To compare our DR results for the proton intermediate state contribution with the data, we perform, for every fixed value of $ Q^2 $, one subtraction for the amplitude $ \cF_3 $ with the subtraction point fixed by one cross section result, which we take from Ref.~\cite{Bernauer:2013tpr}. We like to caution that the two-parameter "empirical" extraction of Ref.~\cite{Bernauer:2013tpr}  is too simplified to be interpreted as "data". In order to obtain an empirical TPE extraction, one would have to apply a full dispersion formalism (elastic + inelastic) and provide a fit of the subtraction function directly to the elastic scattering observables. The present work is a first necessary step towards this aim. Any comparison with the simplified TPE extraction in Ref.~\cite{Bernauer:2013tpr} which we give in the following should therefore only be considered as qualitative.

\begin{figure}[htp]
\begin{center}
\includegraphics[width=.63\textwidth]{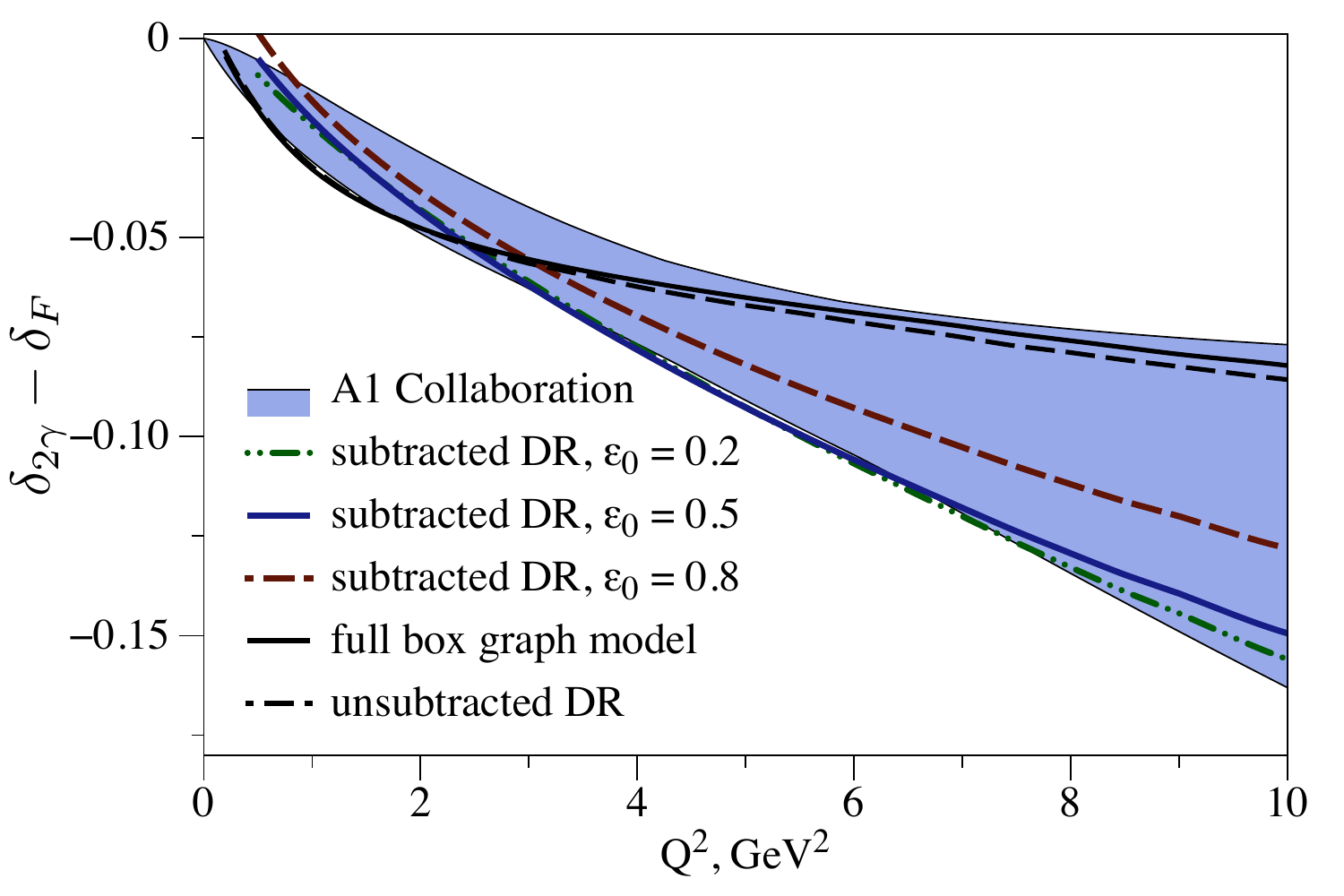}
\end{center}
\caption{ Subtracted DR based prediction for the TPE corrections $ \delta_{2 \gamma} -\delta_F $, in comparison with the box diagram model prediction, unsubtracted DR prediction, for $ \varepsilon = 0.01 $, and with the parametrization of experimental data \cite{Bernauer:2013tpr}, for $ \varepsilon = 0 $ (blue band). The subtracted DR predictions are shown for three choices of the subtraction point: $ \varepsilon_0 = 0.2,~0.5, ~0.8 $.}
\label{delta_vs_Q2_subtracted}
\end{figure}

For comparison, we also show the result for the box diagram model in Fig. \ref{delta_vs_Q2_subtracted}. 
The difference between the results for different choices of the subtraction point corresponds to the uncertainty of our procedure. We would like to notice that for $Q^2$ larger than around $ 1 ~ \mathrm{GeV}^2 $ the account of inelastic intermediate states becomes increasingly important. Also a description in terms of intermediate hadronic states ceases to be valid for large momentum transfer: due to the scattering off individual quarks, one will go over into a partonic picture \cite{Borisyuk:2007re,Kivel:2009eg,Chen:2004tw, Afanasev:2005mp,Kivel:2012vs}.

\begin{figure}[htp]
\begin{center}
\includegraphics[width=\textwidth]{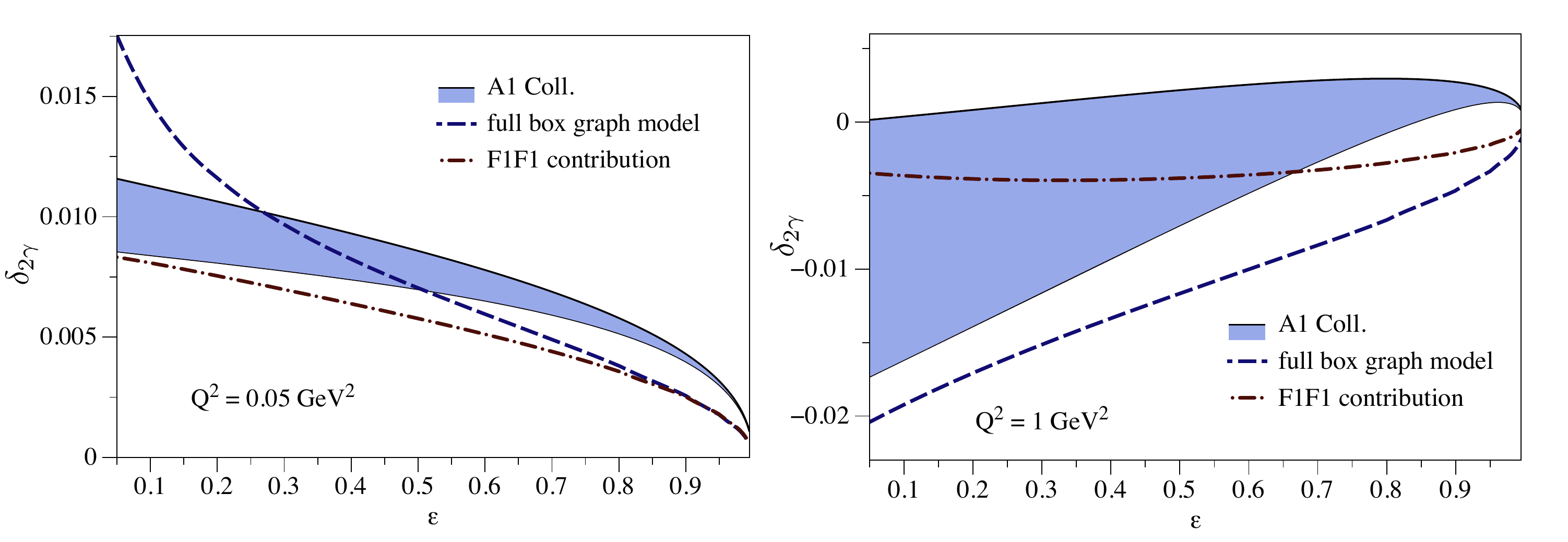}
\end{center}
\caption{ Model prediction for the TPE correction for 
$ Q^2 = 0.05 ~\mathrm{GeV}^2 $ (left panel) and 
$ Q^2 = 1 ~\mathrm{GeV}^2 $ (right panel).
Dashed curve: full box diagram model result; dashed-dotted curve: $ \mathrm{F}_1 \mathrm{F}_1$ vertex contribution only. The experimental results from the MAMI/A1 Coll. \cite{Bernauer:2013tpr} are shown by the blue bands.}
\label{delta_vs_epsilon}
\end{figure}

We next discuss in more detail the TPE evaluations using nucleon intermediate state only in the region of low momentum transfers, to test the validity of this approximation. The TPE correction to the unpolarized elastic electron-proton scattering evaluated in the box diagram model of Section \ref{sec7} is shown in Fig.~\ref{delta_vs_epsilon} as a function of $ \varepsilon $ for momentum transfers $ Q^2 = 0.05 ~\mathrm{GeV}^2 $ and $ Q^2 = 1 ~\mathrm{GeV}^2 $. Our model calculation results are in agreement with a similar calculation performed by Blunden, Melnitchouk and Tjon \cite{Blunden:2003sp}. 
For small momentum transfers, the model calculation approaches the Feshbach limit, and is in agreement with the experimental results.

\begin{figure}[htp]
\begin{center}
\includegraphics[width=\textwidth]{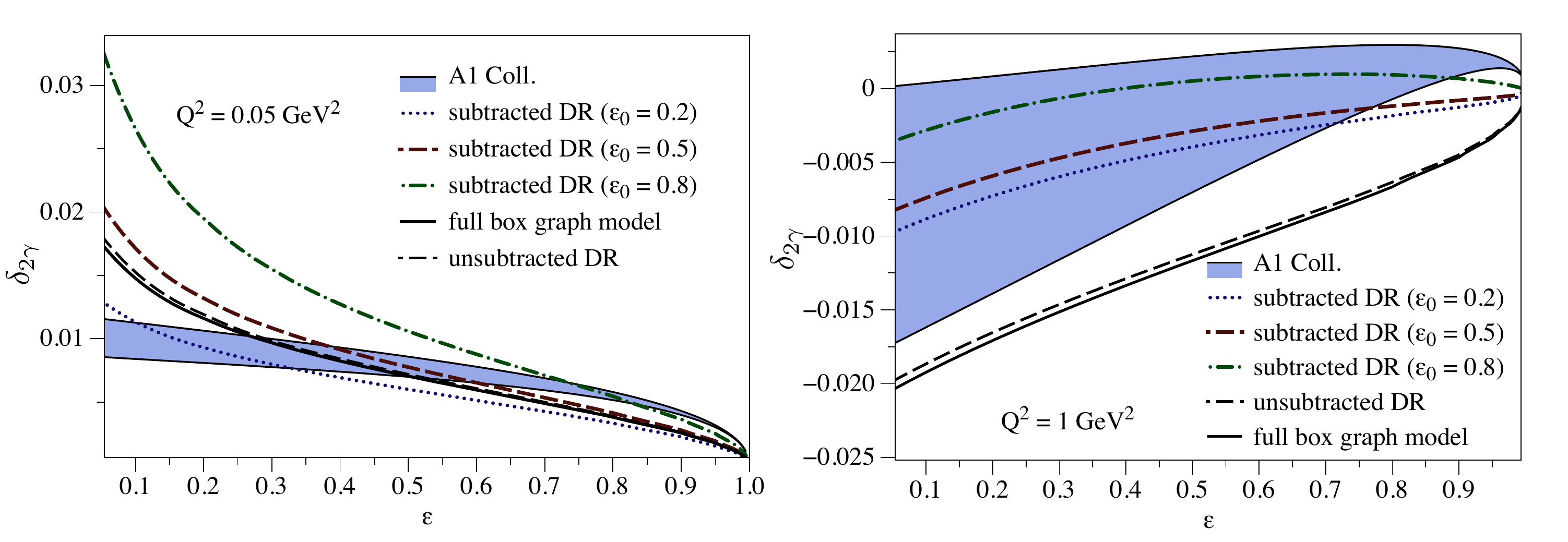}
\end{center}
\caption{Subtracted DR based predictions for the TPE corrections for 
$ Q^2 = 0.05 ~\mathrm{GeV}^2 $ (left panel) and 
$ Q^2 = 1 ~\mathrm{GeV}^2 $ (right panel), in comparison with the unsubtracted DR prediction as well as with the box diagram model. The subtracted DR curves correspond with three choices for the subtraction points~: $ \varepsilon_0 = 0.2,~0.5, ~0.8 $. The blue bands correspond  with the experimental result from the fit of Ref.~\cite{Bernauer:2013tpr}. }
\label{delta_subtracted_plots_vs_epsilon}
\end{figure}

We next show our predictions at low momentum transfers based on the subtracted DR framework. As seen from Fig.~\ref{delta_subtracted_plots_vs_epsilon}, the subtracted DR result describes the data better in the region of intermediate $ \varepsilon $. For higher $ \varepsilon $ values, i.e., higher energies, the contribution of inelastic intermediate states become important and the agreement between theory and experiment becomes worse. One also notices clear deviations at lower values of $ \varepsilon $. This may arise due to the assumption in the experimental TPE analysis of a linear $ \varepsilon $-behavior for the difference $ \delta_{2 \gamma} -\delta_F $. The theoretical calculations show non-linear behaviour in $ \varepsilon $ for this region.

\begin{figure}[htp]
\begin{center}
\includegraphics[width=.65\textwidth]{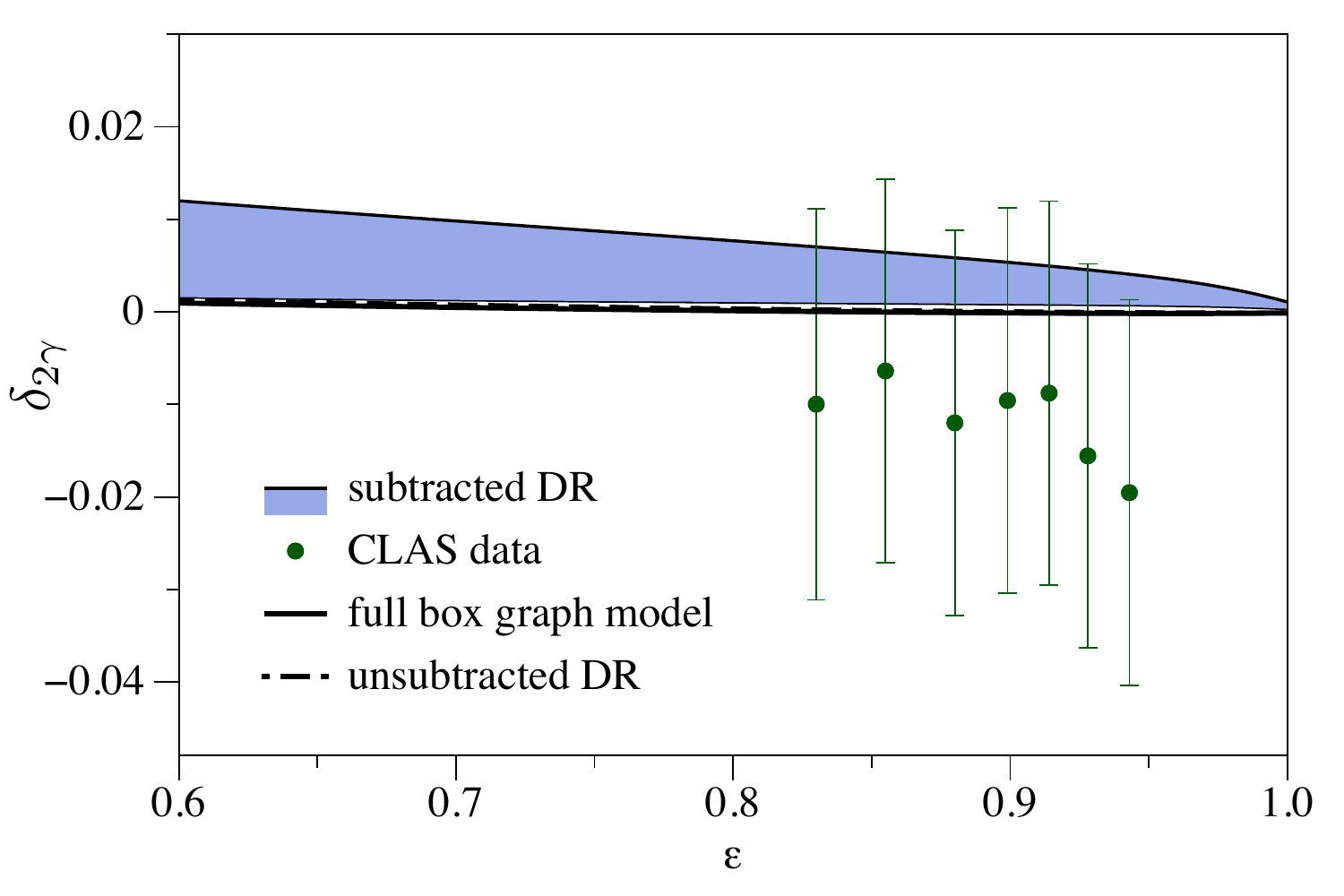}
\end{center}
\caption{Comparison of the subtracted DR prediction for the TPE correction for $ Q^2 = 0.206 ~ \mathrm{GeV}^2 $ with the data \cite{Moteabbed:2013isu}, with the unsubtracted DR prediction and with the box diagram model. The subtraction point used in the DR analysis is $ \varepsilon_0 = 0.83 $.}
\label{delta_CLAS_Zhan}
\end{figure}

\begin{figure}[htp]
\begin{center}
\includegraphics[width=.65\textwidth]{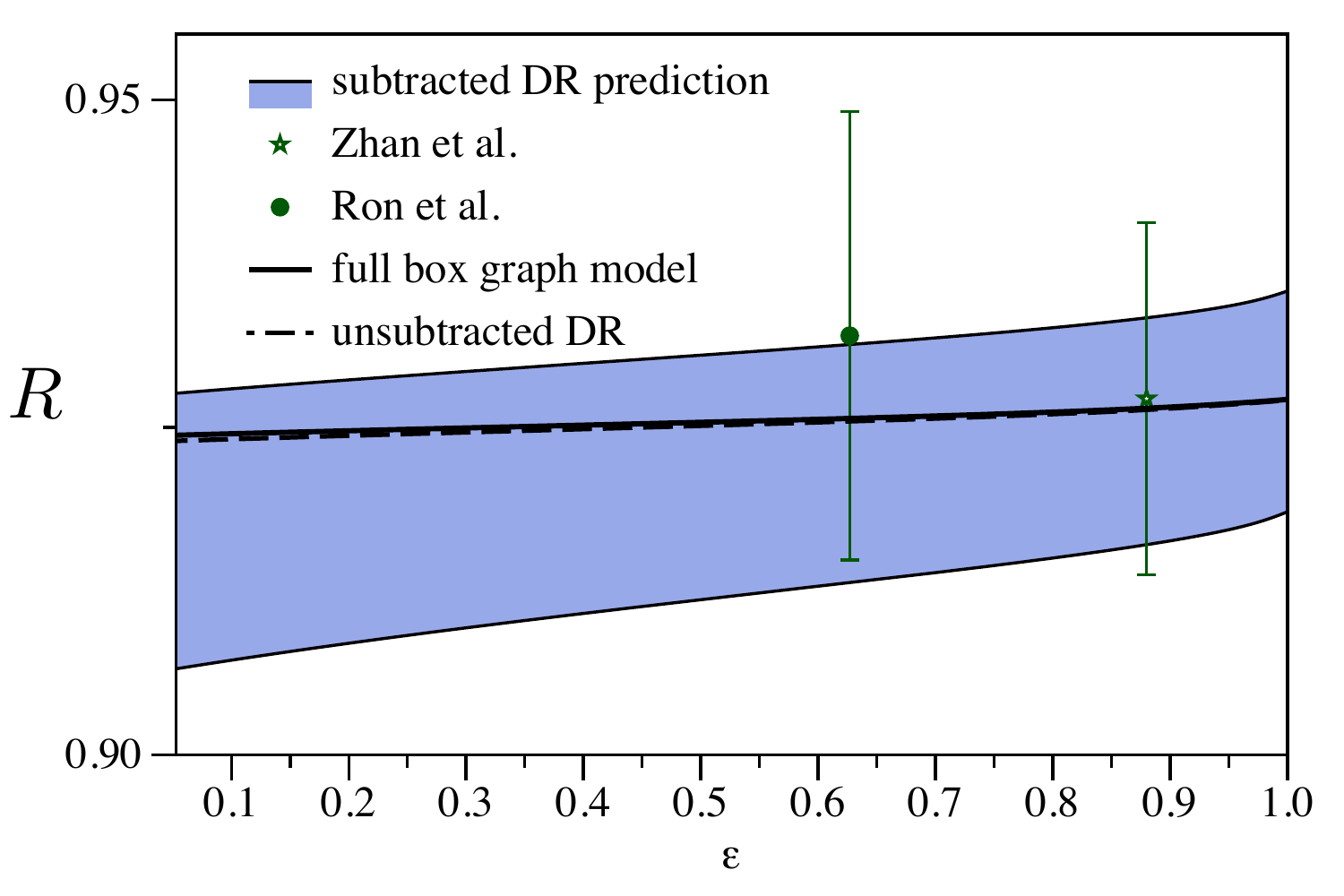}
\end{center}
\caption{Comparison of the subtracted DR prediction for the ratio $ R = -\mu_p \sqrt{\frac{1+\varepsilon}{\varepsilon}\tau} \frac{P_t}{P_l}$  for $ Q^2 = 0.298 ~ \mathrm{GeV}^2 $ with the data \cite{Zhan:2011ji,Ron:2011rd}, with the unsubtracted DR prediction and with the box diagram model. The subtraction point used in the DR analysis is $ \varepsilon_0 = 0.83 $.}
\label{ptpl_CLAS_Zhan}
\end{figure}

For $ Q^2 \approx 0.206 ~\mathrm{GeV}^2 $,  the CLAS Collaboration has recently performed  measurements of the ratio of $ e^{+} p $ to $ e^{-} p $ elastic scattering cross section \cite{Moteabbed:2013isu}. Its deviation from unity is directly related to the TPE corrections. 
Furthermore, the ratio $ P_t / P_l $ was measured for momentum transfer values $ Q^2 = 0.298 ~\mathrm{GeV}^2 $ \cite{Zhan:2011ji} and $ Q^2 = 0.308 ~\mathrm{GeV}^2 $ \cite{Ron:2011rd} in Hall A of JLab. In Figs.~\ref{delta_CLAS_Zhan}, \ref{ptpl_CLAS_Zhan} we show the theoretical estimates for physical observables based on the subtracted DR prediction. We fix the subtracted amplitude $ \cF_3 $ according to Eq. (\ref{subtraction_term}), by using the unpolarized cross section analysis of Ref.~\cite{Bernauer:2013tpr} at one point in $ \varepsilon $ as input. We choose the subtraction point $ \varepsilon_0 = 0.83 $, which is in the $ \varepsilon $-range of both experiments. For both observables we use the FFs from the $ P_t / P_l $ measurement of 
Ref.~\cite{Zhan:2011ji}. We extract the TPE correction $ \delta_{2 \gamma} $ from the CLAS data of the cross section ratio $ R_{2 \gamma} = \sigma(e^{+} p) / \sigma(e^{-} p) $  by $ \delta_{2 \gamma} \approx (1 + \delta_{even}) \times (1-R_{2 \gamma})/2$, where $\delta_{even} \approx -0.2$ 
is the total charge-even radiative correction factor according to Ref. \cite{Moteabbed:2013isu}. Note that for the CLAS data, which have been radiatively corrected according to the Mo and Tsai (MT) procedure~\cite{Mo:1968cg} in Ref. \cite{Moteabbed:2013isu}, we applied the correction $ \delta^{MT}_{2 \gamma, ~soft} - \delta^{MaTj}_{2 \gamma, ~soft}$ to the data in order to compare relative to the Maximon and Tjon (MaTj) procedure which we follow in this paper. The bound on the subtracted DR analysis arises from the experimental uncertainty entering through the subtraction. We conclude from Figs. \ref{delta_CLAS_Zhan}, \ref{ptpl_CLAS_Zhan} that all measurements are in agreement for small momentum transfers and the TPE corrections are described by the elastic contribution within the errors of the experiments.

\begin{figure}[htp]
\begin{center}
\includegraphics[width=\textwidth]{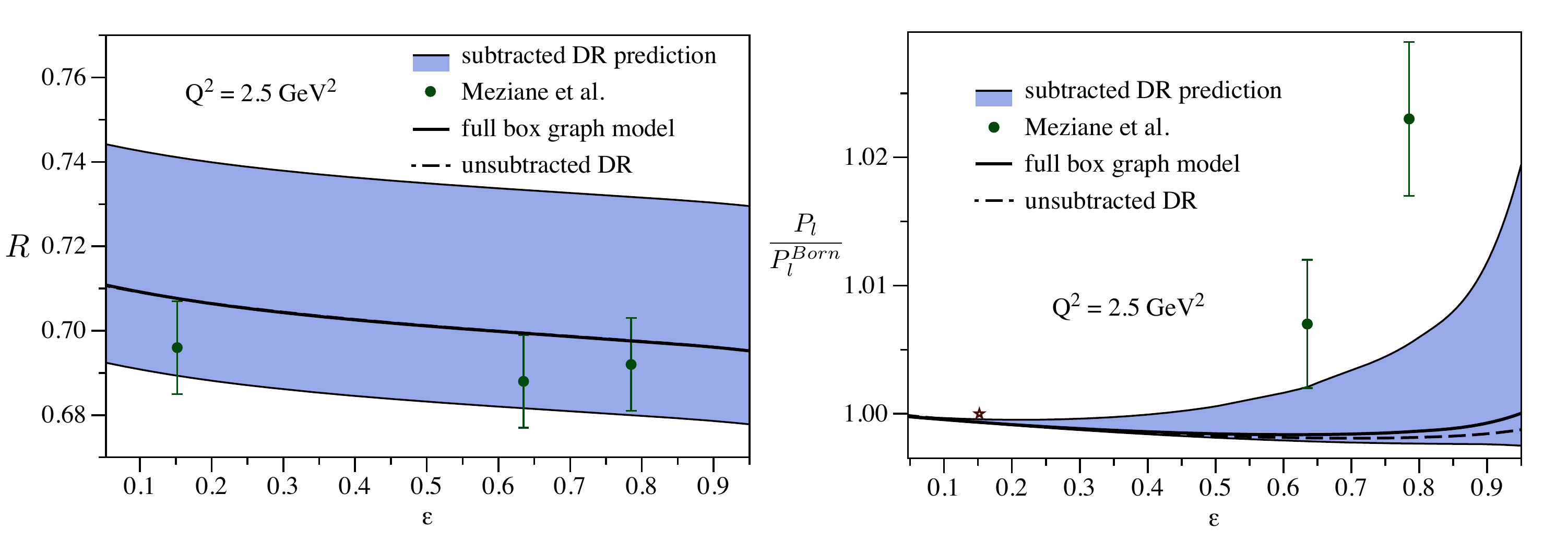}
\end{center}
\caption{Comparison of the subtracted DR predictions for the ratio $ R = -\mu_p \sqrt{\frac{1+\varepsilon}{\varepsilon}\tau} \frac{P_t}{P_l}$ (left panel) and 
$ P_l / P^{Born}_l $ (right panel)  for $ Q^2 = 2.5 ~ \mathrm{GeV}^2 $ with the data 
of Ref.~\cite{Meziane:2010xc}, with the unsubtracted DR prediction and with the box diagram model. 
The subtraction point used in the DR analysis is $ \varepsilon_0 = 0.785 $. }
\label{ptplplpl}
\end{figure}

We next discuss the polarization transfer observables for momentum transfer $ Q^2 \approx 2.5 ~ \mathrm{GeV}^2 $ where data have been taken both for $ P_t $ and $ P_l $ separately \cite{Guttmann:2010au}. In our theoretical predictions, we use the $1 \gamma$-exchange FFs taken from the $ P_t / P_l $ ratio measurement. To evaluate the TPE structure amplitudes, we use the dipole FFs as an input. The comparison with the data for the ratio $ P_t / P_l $ is shown in Fig. \ref{ptplplpl}. As one sees, the present data for $ P_t / P_l $ \cite{Meziane:2010xc} does not allow to extract a TPE effect, indicating a cancellation between the three TPE amplitudes for this specific observable.
The comparison with the data  \cite{Meziane:2010xc} for the absolute polarization transfer observable $ P_l/P^{Born}_l $ \cite{Meziane:2010xc} is also shown in Fig. \ref{ptplplpl}. 
It shows that the point at $ \varepsilon = 0.635 $ with $ P_l/P^{Born}_l = 1.007 \pm 0.005 $ is consistent with the proton contribution only, but the point at $ \varepsilon = 0.785 $ with $ P_l/P^{Born}_l = 1.023 \pm 0.006 $ requires further theoretical investigations, e.g., account of inelastic intermediate states which are relevant at these larger momentum transfers. The specific property of the subtracted DR analysis for the ratio $ P_l/P^{Born}_l $ is the divergence of the errors for $ \varepsilon \to 1 $ as $ 1/\sqrt{1-\varepsilon}$.

\section{Conclusions and outlook}
\label{sec9}
In this work we have studied the TPE corrections to elastic electron-proton scattering 
with the aim to minimize the model dependence when applied to data. 
For this purpose we have studied a subtracted dispersion relation formalism where the 
real part of the $ \cF_3 $ structure amplitude is reconstructed from the 
corresponding imaginary parts through a subtracted dispersion relation. We have related 
the subtraction constant at a fixed value of $Q^2$ to a precisely measured cross section 
point at one value of $\varepsilon$. The remaining $\varepsilon$ dependence of the cross section, as well as the other observables then follow as predictions in our formalism. 
In this work, we have tested this formalism on the elastic, i.e. proton intermediate state, TPE contribution. We have made a detailed comparison with existing data. 
In the low momentum transfer region, where the nucleon intermediate state contribution is expected to dominate, the presented formalism provides a flexible framework to provide a more accurate extraction of the TPE correction to elastic electron-nucleon scattering. 
At larger values of $Q^2$, the presented subtracted dispersion relation formalism can be extended in a next step to include inelastic intermediate state contributions. 
Moreover, a further extension of the subtracted DR formalism is to evaluate the TPE corrections for the case of muon-proton scattering at low energies, which requires the inclusion of lepton-mass correction terms. A first step in this direction was already performed~\cite{Tomalak:2014dja}.  

\section*{Acknowledgements}
We thank J. Bernauer for providing us with the details of the analysis of experimental data on TPE, as well as C.E. Carlson, M. Gorchtein, N. Kivel, and V. Pascalutsa for technical support and useful discussions. This work was supported by the Deutsche Forschungsgemeinschaft DFG in part through the Collaborative Research Center [The Low-Energy Frontier of the Standard Model (SFB 1044)], in part through the Graduate School [Symmetry Breaking in Fundamental Interactions (DFG/GRK 1581)],
and in part through the Cluster of Excellence [Precision Physics, Fundamental
Interactions and Structure of Matter (PRISMA)].

\appendix

\section{Phases entering the unitarity relations phases}
\label{app1}

The unitarity relation phases entering Eq. (\ref{unitarity_im}) can be expressed in terms of the Mandelstam variables as
\ber \label{angles}
 \cos \phi' & = & \frac{1}{\sqrt{4 Q^2 Q^2_1 x x_1} x_2} \left( -Q^2_2 + Q^2 x + Q^2_1 x_1 + \frac{s Q^2_2}{(s-M^2)^2} Q^2_2 (x Q^2_1 + x_1 Q^2 ) \right), \nonumber  \\
 \cos \tilde{\phi} & = & \frac{1 }{ \sqrt{4 Q^2 Q^2_2 x x_2 } }  \left( x Q^2_2 + x_2 Q^2 - Q^2_1 \right),  \nonumber  \\
 \cos \tilde{\phi}' & = & \frac{ 1 }{ \sqrt{4 Q^2_1 Q^2_2 x_1 x_2} }  \left( - Q^2 + x_2 Q^2_1 + x_1 Q^2_2 \right), \nonumber \\
 \cos (\phi - \phi') & = & \frac{ 1  }{2 x x_1 x_2}    \left( x^2 + x_1^2 + x_2^2 - 1 + 2 \frac{s^3}{(s-M^2)^6} Q^2 Q^2_1 Q^2_2 \right),  \nonumber  \\
 \cos (\phi + \tilde{\phi}) & = &  \frac{ 1 }{ \sqrt{4 x_1 x_2 Q^2_1 Q^2_2 } } \frac{1}{x} \left( - Q^2_1 x_1  - Q^2_2 x_2  + Q^2 - \frac{s Q^2}{(s-M^2)^2} (Q^2_2 + Q^2_1) x \right),   
\eer
with 
\ber
 x & \equiv & \frac{1}{2}( 1 + \cos \cma) = 1 - \frac{s Q^2}{(s-M^2)^2}, \nonumber \\
 x_1 & \equiv & \frac{1}{2}( 1 + \cos \theta_1) = 1 - \frac{s Q^2_1}{(s-M^2)^2} , \nonumber \\
 x_2 & \equiv & \frac{1}{2}( 1 + \cos \theta_2) = 1 - \frac{s Q^2_2}{(s-M^2)^2}. 
 \eer

\section{Different integration coordinates in unitarity relations}
\label{app2}

The boundaries of the ellipse mentioned in Section \ref{sec6} correspond to $ \cos^2 \phi_1 = 0 $. Defining $ z_1 \equiv  \cos \theta_1, ~z_2 \equiv  \cos \theta_2, ~z \equiv  \cos \cma $ the ellipse equation is given by 
\ber
 1 - z^2 - z_1^2 - z^2_2 & = & - 2 z z_1 z_2,
\eer
The coordinates $ z_1, z_2 $ can be rotated by $ 45^0 $, so that the new coordinate system coincides with the axes of the ellipse
\ber
 \tilde{z}_1 & = & - \frac{1}{\sqrt{2}} ( z_1 + z_2 ), ~~~~ \tilde{z}_2 =  \frac{1}{\sqrt{2}} ( z_1 - z_2 ).
\eer
The $ \tilde{z}_2 $-axis corresponds to the line $ Q^2_1 = Q_2^2 $, whereas the $ \tilde{z}_1 $-axis corresponds to the line $ Q^2_2 = Q_{max}^2- Q_1^2 $. The phase space integration in terms of new coordinates is expressed as
\ber \label{appendix_int}
 \mathop{\mathlarger{\int}}  \mathrm{d} \Omega & = & 2 \mathop{\mathlarger{\int}} \limits^{~~1}_{-1}  \mathrm{d} \cos \theta_1  \mathop{\mathlarger{\int}} \limits^{~~\pi}_0  \mathrm{d} \phi_1 = \frac{2}{\sqrt{1-z^2} } \mathop{\mathlarger{\int}}  \mathrm{d} \tilde{z}_1  \mathrm{d} \tilde{z}_2 \frac{1}{|\alpha|},
\eer
with $ \alpha \equiv \sin \theta_1 \sin \phi_1$. The ellipse equation is then given by
\ber
 \frac{\tilde{z}^2_1}{1+z} + \frac{\tilde{z}^2_2}{1-z} & = & 1.
\eer
The integration of Eq. \ref{appendix_int} maps out the whole surface of the ellipse. It is therefore convenient to introduce the elliptic coordinates $ \alpha, \phi $ as
\ber
 \tilde{z}_1 & = & \sqrt{1 - \alpha^2}   \sqrt{1 + z} \cos (\phi), \nonumber \\
 \tilde{z}_2 & = & \sqrt{1 - \alpha^2}   \sqrt{1 - z} \sin (\phi). 
\eer
which satisfy
\ber
 \frac{\tilde{z}^2_1}{1+z} + \frac{\tilde{z}^2_2}{1-z} & = & 1 - \alpha^2.
\eer
The photons virtualities  $ Q^2_1 $, $ Q^2_2 $ are symmetric in terms of the elliptic coordinates $ \alpha, \phi $. The phase space integration in terms of these elliptic coordinates can then be expressed as
\ber
 \mathop{\mathlarger{\int}}  \mathrm{d} \Omega & = & \frac{2}{\sqrt{1-z^2} } \mathop{\mathlarger{\int}}  \mathrm{d} \tilde{z}_1  \mathrm{d} \tilde{z}_2 \frac{1}{|\alpha|} = 2  \mathop{\mathlarger{\int}} \limits^{~~1}_{0}  \mathrm{d} \alpha  \mathop{\mathlarger{\int}} \limits^{~~2\pi}_0  \mathrm{d} \phi.
\eer

\section{Box diagram results in terms of LOOPTOOLS integrals}
\label{app3}

We describe the details of the box diagram calculation for the point-like model below. The helicity amplitude from the direct and crossed TPE diagram is given by Eq. (\ref{helamp}), and has the following structure
\ber
& T_{dir} = N^0 A_{dir} +  N_\alpha A^\alpha_{dir} + N_{\alpha \beta} A^{\alpha \beta}_{dir} + N_{\alpha \beta \gamma} A^{\alpha \beta \gamma}_{dir} + N_{\alpha \beta \gamma \delta} A^{\alpha \beta \gamma \delta}_{dir} ,
\eer
\ber
 & & (A_{dir}, A^\alpha_{dir}, A^{\alpha \beta}_{dir}, A^{\alpha \beta \gamma}_{dir},A^{\alpha \beta \gamma \delta}_{dir}) = i \mathop{\mathlarger{\int}} \frac{ \mathrm{d}^4 k_1}{( 2 \pi )^4} \nonumber \\
 & & \frac{( 1 , k^\alpha_1 , k^\alpha_1 k^\beta_1 , k^\alpha_1 k^\beta_1 k^\gamma_1, k^\alpha_1 k^\beta_1 k^\gamma_1k^\delta_1  )}{ ((k_1 - P - K )^2 - M^2)(k_1^2 - m^2)((k_1 - K - \frac{q}{2} )^2 - \mu^2 )((k_1 - K + \frac{q}{2} )^2 - \mu^2 )},
\eer
with $ N $ - spinor contraction with free indices. The contraction is done with momentums from the expansion of the integrals $ A_{dir}, A^\alpha_{dir}, ... $ in terms of the on-shell momentums. These integrals are invariant under the replacement $ q \to -q $ and can be expressed as
\ber
 A^\alpha_{dir} & = & a_s ( P + K )^\alpha + a_P P^\alpha, \nonumber \\
 A^{\alpha \beta}_{dir} & = & a_{ss} ( P + K )^{\alpha} ( P + K )^{\beta} + a_{sP} P^{[\alpha,} ( P + K )^{\beta]} + a_{PP} P^\alpha P^\beta + a_{qq} q^\alpha q^\beta + a_{00} g^{ \alpha \beta }, \nonumber \\
 A^{\alpha \beta \gamma}_{dir} & = & a_{sss} ( P + K )^\alpha ( P + K )^\beta ( P + K )^\gamma + a_{PPP} P^\alpha P^\beta P^\gamma + a_{s00} g^{[\alpha, \beta,}  ( P + K )^{\gamma]} \nonumber \\
& + & a_{ssP} P^{[\alpha,} ( P + K )^{\beta,} ( P + K )^{\gamma]} + a_{sPP} P^{[\alpha,} P^{\beta,} ( P + K )^{\gamma]} + a_{Pqq} q^{[\alpha,} q^{\beta,} P^{\gamma]}  \nonumber \\
& + & a_{P00} g^{[\alpha ,\beta, } P^{\gamma]}  + a_{sqq} q^{[\alpha,} q^{\beta,} ( P + K )^{\gamma]}, \nonumber \\
 A^{\alpha \beta \gamma \delta}_{dir} & = & a_{ssss} ( P + K )^\alpha ( P + K )^\beta ( P + K )^\gamma  ( P + K )^\delta + a_{PPPP} P^\alpha P^\beta P^\gamma P^\delta + a_{qqqq} q^\alpha q^\beta q^\gamma q^\delta  \nonumber \\
& + & a_{sssP} P^{[\alpha,} ( P + K )^{\beta,} ( P + K )^{\gamma,} ( P + K )^{\delta]} + a_{sPPP} P^{[\alpha,} P^{\beta,} P^{\gamma,} ( P + K )^{\delta]}  \nonumber \\
& + & a_{ssPP}  P^{[\alpha,} P^{\beta,} (P+K)^{\gamma,} (P+K)^{\delta]} + a_{0000} g^{[\alpha, \beta,} g^{\gamma, \delta]} + a_{sP00} g^{[\alpha, \beta,} (P+K)^{\gamma,}P^{\delta]} \nonumber \\
& + &  a_{PP00} g^{[\alpha, \beta,} P^{\gamma,}P^{\delta]} + a_{qq00} g^{[\alpha, \beta,} q^{\gamma,}q^{\delta]} + a_{ss00} g^{[\alpha, \beta,} (P+K)^{\gamma,}(P+K)^{\delta]}  \nonumber \\
& + & a_{sPqq} (P+K)^{[\alpha,} P^{\beta,} q^{\gamma,}q^{\delta]} + a_{PPqq} q^{[\alpha,} q^{\beta,} P^{\gamma,} P^{\delta]} + a_{ssqq} q^{[\alpha,} q^{\beta,} (P+K)^{\gamma,} (P+K)^{\delta]},
\eer
where all non-equivalent permutations are only accounted once. The integrals $ a_s, a_p, ...$can be expressed through the LOOPTOOLS four-point functions for kinematics
\ber
& m_1 = m_e,~~ m_2 = \mu, ~~ m_3 = M, ~~ m_4 = \mu, \nonumber \\
& p_1 = - k', ~~ p_2 = - p', ~~ p_3 = p, ~~ p_4 = k, \nonumber \\
& p_{12} \equiv (p_1+p_2)^2 = s,~~ p_{23} \equiv (p_2+p_3)^2 = t, \nonumber \\
& k_1 = -k',~~ k_2 = - ( P + K ), ~~ k_3 = - ( P + K ) + p = -k. 
\eer

The result for the $ \mathrm{F}_1 \mathrm{F}_1 $ vertex structure of virtual photon-proton-proton vertices for the direct diagram is given by
\ber
 \cG_M & = & -e^2(2(-M^2 t + (s-M^2)^2)a_p + 2 ((s-M^2)^2- t s) a_s -  4 (s-M^2)(M^2-\frac{t}{4}) a_{pp}   \nonumber \\
 & - & 2 (s^2-M^4 - t s) a_{ss} - 12 (s-M^2-\frac{t}{3})a_{00} - 2((s-M^2)(s+3 M^2) - t s) a_{sp}\nonumber \\
 & - & 4 (s - M^2) t a_{qq}), \nonumber \\
 \cF_2 & = & 2 M^2 e^2 t (a_{pp} + a_{sp}), \nonumber \\
 \cF_3 & = & -e^2 (-4 (s-M^2) M^2 a_p -4 (s-M^2)M^2 a_s + 8 M^2 (M^2 -\frac{t}{4}) a_{pp} \nonumber \\
 & + &  4(s+M^2) M^2 a_{ss} + 24 M^2 a_{00} + 4(s+3M^2) M^2 a_{sp} + 8 M^2 t a_{qq}).
\eer
The result for the $ \mathrm{F}_1 \mathrm{F}_2 $ vertex structure for the direct diagram is given by
\ber
 \cG_M & = & e^2( \left(4 M^2-t\right) \left( s - M^2 + \frac{t}{2}\right) a_{PPP} + 4 s \left(s-M^2\right) a_{sss} + 4 t \left(s-M^2\right) a_{sqq}    \nonumber \\
 & + & (-8 M^4+M^2 (4 s+3 t)+s (4 s+t)) a_{sPP} + 24 \left(s-M^2\right) a_{s0} + 12 \left(2 (s - M^2)+t\right) a_{P0}  \nonumber \\ 
 & + & 2 t \left(2 (s - M^2)+t\right) a_{Pqq} + (-4 M^2(s+M^2) + 2 s (4 s + t)) a_{ssP})   \nonumber \\
 & + &  e^2( ( s+ M^2) t a_p + 2  t s a_s - ( 4 ((s-M^2)^2 + s t) - t (M^2 - \frac{t}{4})) a_{pp}  \nonumber \\
& - & (4 (s-M^2)^2 + 3 s t) a_{ss} + ( 8 (s - M^2 ) - 4 t) a_{00} - (8(s-M^2)^2 - 7 t s + M^2 t) a_{sp},   \nonumber \\
 \cF_2 & = & e^2( (s+M^2) t a_p + 2 s t a_s -( ( 2s + M^2 ) t - \frac{t^2}{4})  a_{pp}  - 3 s t a_{ss} - 8 t a_{00} - 6 t s a_{sp} - t^2 a_{qq} )   \nonumber \\
 & + &  e^2(  t \left(3 M^2- \frac{3}{4} t\right) a_{PPP} + s t a_{sss} + t \left(4 s + M^2\right) a_{ssP} + t ( 3 s + 4 M^2 -\frac{t}{4}) a_{sPP} \nonumber \\
  & + &  6t a_{s0} + 18 t a_{P0} + 3 t^2 a_{Pqq} + t^2 a_{sqq} ),  \nonumber \\
 \cF_3 & = & 2 M^2 e^2(  2 \left(2 (s - M^2 )+t\right)  a_{pp} + 4 (s - M^2) a_{ss} - 8  a_{00} +  (8 ( s - M^2 ) +t)  a_{sp} - 4 t a_{qq} )  \nonumber \\
  & + & 2 M^2 e^2( ( t-4 M^2) a_{PPP} -  4 s a_{sss} - 4 \left(2 s + M^2\right) a_{ssP} + \left(-8 M^2-4 s+t\right) a_{sPP} \nonumber \\
  & - &  24 a_{s0} - 24 a_{P0} - 4 t a_{Pqq} - 4 t a_{sqq} ).
\eer
The result for the $ \mathrm{F}_2 \mathrm{F}_2 $ vertex structure for the direct diagram is given by
\ber
 \cG_M & = & e^2 (\frac{9 \left(-4 M^2+4 s+t\right)}{2 M^2}a_{0000} + \frac{t \left(-44 M^2+44 s+19 t\right)}{4 M^2} a_{qq00} \nonumber \\
 & - & \frac{\left(4 M^2-t\right) \left(44 M^2-44 s-19 t\right)}{16 M^2} a_{PP00}  + \frac{1}{2} \left(-M^4+\frac{2 s^3}{M^2}-4 M^2 s+s (3 s+t)\right) a_{sssP}   \nonumber \\
 & + & (-4 M^2 - 3 s + (7 s^2 + (3 s t)/4)/M^2) a_{SS00} + (\frac{7 s^2+\frac{3 s t}{4}}{M^2}-15 M^2+8 s+\frac{19 t}{4}) a_{sP00}   \nonumber \\
  & - & \frac{\left(t-4 M^2\right)^2 \left(2 M^2-2 s-t\right)}{32 M^2} a_{PPPP} + \frac{t^2 \left(-2 M^2+2 s+t\right)}{2 M^2} a_{qqqq} + \frac{s^3-M^4 s}{2 M^2} a_{ssss} \nonumber \\
  & - & \frac{\left(4 M^2-t\right) \left(5 M^4-2 M^2 (s+t)-s (3 s+t)\right)}{8 M^2} a_{sPPP}  \nonumber \\
 & - &  \frac{t \left(4 M^2-t\right) \left(2 M^2-2 s-t\right)}{4 M^2} a_{PPqq} +  \frac{t \left(-5 M^4+2 M^2 (s+t)+s (3 s+t)\right)}{2 M^2} a_{sPqq}   \nonumber \\
 & + & \frac{t \left(-M^4-2 M^2 s+s (3 s+t)\right)}{2 M^2} a_{ssqq}   \nonumber \\
 & + &   \frac{-16 M^6+M^4 (5 t-12 s)+2 M^2 s (12 s+5 t)+s (s-t) (4 s+t)}{8 M^2} a_{ssPP})  \nonumber \\
 & + &  e^2( -\frac{\left(4 M^2-t\right) \left(3 M^4-M^2 (6 s+t)+s (3 s+2 t)\right)}{8 M^2} a_{PPP}  \nonumber \\
& + & \frac{\left(-28 M^6+M^4 (48 s+9 t)-2 M^2 s (6 s+11 t)+s \left(-8 s^2+s t+3 t^2\right)\right)}{8 M^2} a_{sPP}  \nonumber \\
  & + & \frac{\left(-4 M^6+3 M^4 s+M^2 s (6 s-t)-s^2 (5 s+2 t)\right)}{2 M^2}a_{ssP}  -\frac{ \left(M^2-s\right) \left(9 M^2-7 s-4t\right)}{M^2}a_{P00}   \nonumber \\
  & - & \frac{ \left(13 M^4-24 M^2 s+11 s^2+5 s t\right)}{M^2} a_{s00} - \frac{s \left(3 M^4-6 M^2 s+s (3 s+t)\right)}{2 M^2}a_{sss}  \nonumber \\ 
  & + & \frac{t \left(-3 M^4+M^2 (6 s+t)-s (3 s+2 t)\right)}{2 M^2}a_{Pqq} -\frac{t  \left(5 M^4-10 M^2 s+s (5 s+3 t)\right)}{2 M^2} a_{sqq} )   \nonumber \\
 & + &  e^2( \frac{\left(s t \left(M^2-s\right)+\left(M^2-s\right)^3\right)}{2 M^2}a_p - \frac{t^2 \left(M^2+s\right)-2 t \left(M^2-s\right)^2+4 \left(M^2-s\right)^3}{8 M^2}a_{pp}  \nonumber \\
  & - & \frac{\left(M^2-s\right) \left(3 M^4-6 M^2 s+s (3 s+2 t)\right)}{2 M^2} a_{ss}  + \frac{ \left(s t \left(M^2-s\right)+\left(M^2-s\right)^3\right)}{2 M^2} a_s  \nonumber \\
  & - & \frac{\left(M^2-s\right) \left(4 M^4-M^2 (8 s+t)+s (4 s+3 t)\right)}{2 M^2} a_{sp} + (\frac{s (4 s+3 t)}{M^2}+4 M^2-8 s-t) a_{00},   \nonumber \\
 \cF_2 & = & e^2 ( 6 t a_{PP00} + 2 t a_{SS00} + 8 t a_{sP00} + \frac{s t}{4} a_{ssss} + \frac{3}{16} t \left(4 M^2-t\right) a_{PPPP}   \nonumber \\
 & + &  \frac{1}{4} t \left(M^2+5 s\right) a_{sssP} + \frac{1}{4} t \left(7 M^2+3 s-t\right) a_{sPPP} +  t^2 a_{sPqq} + \frac{3 t^2}{4} a_{PPqq} + \frac{t^2}{4} a_{ssqq}  \nonumber \\
 & + &    \frac{1}{16} t \left(20 M^2+28 s-t\right) a_{ssPP})  \nonumber \\
 & + &  e^2( -\frac{1}{8} t  (4 s+t) a_{PPP} - \frac{1}{2} s t  a_{sss} - \frac{1}{2} t \left(M^2+2 s\right) a_{ssP} +  \frac{1}{8} t \left(-4 M^2-8 s+t\right) a_{sPP}  \nonumber \\
 & - &  3 ta_{s00} + 2 t a_{P00} + \frac{t^2}{2} a_{Pqq} -\frac{t^2 }{2} a_{sqq} )   \nonumber \\
 & + &  e^2( \frac{1}{4} t \left(2 M^2-2 s-t\right)  a_{pp} +t^2 a_{qq} + 2 t a_{00} + \frac{1}{2} t \left(M^2-s\right) a_{sp} ),\nonumber
 \eer
 \ber
 \cF_3 & = & e^2 (-36 a_{0000} - 22 t a_{qq00} + \frac{11}{2} \left(t-4 M^2\right) a_{PP00} -2 (4 M^2 + 7 s + t) a_{SS00}  \nonumber \\
 & - &  2 (15 M^2 + 7 s + t) a_{sP00} - \frac{1}{4} s \left(4 \left(M^2+s\right)+t\right) a_{ssss} - \frac{1}{8} \left(t-4 M^2\right)^2 a_{PPPP} -2 t^2 a_{qqqq} \nonumber \\
  & + & (-M^4-\frac{1}{4} M^2 (20 s+t)-\frac{1}{2} s (4 s+t)) a_{sssP} -\frac{1}{16} \left(4 M^2-t\right) \left(20 M^2+12 s+t\right) a_{sPPP}  \nonumber \\
 & - &  \frac{1}{4} t \left(20 M^2+12 s+t\right)  a_{sPqq} + t \left(t-4 M^2\right) a_{PPqq} -\frac{1}{4} t \left(4 M^2+12 s+t\right) a_{ssqq}  \nonumber \\
  & + &  \frac{1}{16} \left(-64 M^4-4 M^2 (28 s+t)-16 s^2+8 s t+t^2\right)a_{ssPP})  \nonumber \\
  & + & e^2( -\frac{1}{16} \left(4 M^2-t\right) \left(12 M^2-12 s-5 t\right) a_{PPP} +  \left(-3 M^2 s+3 s^2+\frac{5 s t}{4}\right) a_{sss}  \nonumber \\
 & + &  \left(-4 M^4+M^2 \left(\frac{5 t}{4}-s\right)+\frac{5}{2} s (2 s+t)\right) a_{ssP}  + \left(-3 M^2 t+3 s t+\frac{5 t^2}{4}\right) a_{Pqq}  \nonumber \\
 & + &  \left(-26 M^2+22 s+\frac{15 t}{2}\right) a_{s00} + \left(-18 M^2+14 s+\frac{15 t}{2}\right) a_{P00}  \nonumber \\
 & + &   \left(-7 M^4+\frac{5}{4} M^2 (4 s+3 t)+2 s^2-\frac{5 t^2}{16}\right) a_{sPP} + \frac{5}{4} t \left(-4
   M^2+4 s+t\right) a_{sqq} )   \nonumber \\
 & + &  e^2( \left(\left(M^2-s\right)^2+s t\right) a_p + \left(\left(M^2-s\right)^2+s t\right) a_s + \nonumber \\
 & + &   \left(\frac{1}{4} t \left(-2 M^2+2 s+t\right)-\left(M^2-s\right)^2-s t\right) a_{pp} + \left(-4 M^4+8 M^2 s-s (4 s+3 t)\right) a_{sp}   \nonumber \\
 & + &  \left(-3 \left(M^2-s\right)^2-2 s t\right) a_{ss} +  \left(8 M^2-8 s-6 t\right) a_{00} + t \left(2 M^2-2 s-t\right) a_{qq} .
\eer

The crossed diagram contribution to the structure amplitudes can be obtained from the direct diagram contribution by the replacement $ s \to u $, and with a sign according to crossing relations of Eq. (\ref{crossing_relations}).

For the sum of the direct and crossed diagrams the UV divergent terms in the case of $ \mathrm{F}_2 \mathrm{F}_2 $ vertex structure have following behavior
\ber \label{UV_ampl}
 \cG^{UV}_M \sim - \frac{\nu}{ M^2} , ~~~~~~ \cF^{UV}_3 \sim 1  ,   ~~~~~  \cF^{UV}_2 & \sim & 0,
\eer
whereas the amplitudes $ \cF_2, \cG_1, \cG_2 $ are UV finite.

We now describe the details of the calculation for the dipole form of electric and magnetic FFs , see Eq. (\ref{dipole_model}). The Pauli and Dirac FFs have the following expressions
\ber
 F_2 & = & - \frac{(\mu_P - 1) \Lambda^4 4 M^2}{(q^2 - \Lambda^2)^2 (q^2 - 4 M^2)},  \nonumber \\
 F_1 & = &  \frac{\mu_P \Lambda^4}{(q^2 - \Lambda^2)^2} + \frac{(\mu_P - 1) \Lambda^4 4 M^2}{(q^2 - \Lambda^2)^2 (q^2 - 4 M^2)}.
\eer
The amplitudes for the case of dipole form of electromagnetic FFs can be obtained from the point-like model expressions after differentiation of the photon propagators with respect to the IR parameter $ \mu^2 $ and replacement of this parameter either by $ \Lambda^2 $ or by $ 4 M^2 $. The two terms with different order of vector and tensor coupling vertices for the case of the $ \mathrm{F}_1 \mathrm{F}_2 $  vertex structure are not the same in this calculation. Also the tensor expressions for the integrals $ A $ are not symmetric under the replacement $ q \to - q $ in case of both terms for the $ \mathrm{F}_1 \mathrm{F}_2 $ vertex structure. We perform the Passarino-Veltman decomposition \cite{Passarino:1978jh} in terms of the LOOPTOOLS momentums in this case.

The UV divergencies are absent in the calculation with the dipole FFs. We subtract the IR divergencies according to the Maximon and Tjon prescription \cite{Maximon:2000hm}. The soft photons contribution can be easily obtained from the spinor contractions in the numerator of Eq. (\ref{helamp}). If the virtuality of one of the photons $ q_i \to 0$ in case of the $ \mathrm{F}_1 \mathrm{F}_1 $ vertex structure the numerator of the TPE amplitude has a factor $  4 (k p) \bar{u} \gamma_\nu u  \bar{N} \gamma^\nu N $. The IR contribution has to be multiplied by $ F_1(0) F_1(t) $.  The IR contributions to $ \cG_M $ and $ \cF_2 $ structure amplitudes in case of the $ \mathrm{F}_1 \mathrm{F}_2 $ vertex structure has to be multiplied by $ F_1(0) F_2(t) $. In case of the $ \mathrm{F}_2 \mathrm{F}_2 $ vertex structure, the IR divergence is absent.

\end{document}